\documentclass[useAMS,usenatbib,final]{mn2e}

\usepackage{epsfig,graphics}
\usepackage{epsf,graphicx}
\usepackage{color}
\usepackage{xspace}
\usepackage{times}
\usepackage{amssymb,amsmath}

\newcommand{\Ha}{\ifmmode {\rm H}\alpha \else H$\alpha$\fi\xspace}
\newcommand{\Hb}{\ifmmode {\rm H}\beta \else H$\beta$\fi\xspace}
\newcommand{\Hg}{\ifmmode {\rm H}\gamma \else H$\gamma$\gvfi\xspace}
\newcommand{\Hd}{\ifmmode {\rm H}\delta \else H$\delta$\fi\xspace}

\newcommand{\hii}{H~{\sc ii}\xspace}
\newcommand{\heii}{He~{\sc ii}\xspace}

\newcommand{\Hii}{\ifmmode \rm{H}\,\textsc{ii} \else H~{\sc ii}\fi}
\newcommand{\Nii}{[N~{\sc ii}]$\lambda$6584}
\newcommand{\nii}{\ifmmode [\rm{N}\,\textsc{ii}] \else [N~{\sc ii}]\fi}

\newcommand{\oi}{\ifmmode [\rm{O}\,\textsc{i}] \else [O~{\sc i}]\fi}

\newcommand{\neiii}{\ifmmode [\rm{Ne}\,\textsc{iii}] \else [Ne~{\sc iii}]\fi}
\newcommand{\hei}{\ifmmode [\rm{He}\,\textsc{i}] \else [He~{\sc i}]\fi}
\newcommand{\oii}{\ifmmode [\rm{O}\,\textsc{ii}] \else [O~{\sc ii}]\fi}
\newcommand{\Oiii}{[O~{\sc iii}]$\lambda$5007}
\newcommand{\Oiiit}{[O~{\sc iii}]$\lambda$4363}
\newcommand{\oiii}{\ifmmode [\rm{O}\,\textsc{iii}] \else [O~{\sc iii}]\fi}

\newcommand{\sii}{\ifmmode [\rm{S}\,\textsc{ii}] \else [S~{\sc ii}]\fi}
\newcommand{\siii}{\ifmmode [\rm{S}\,\textsc{iii}] \else [S~{\sc iii}]\fi}

\newcommand{\msun}{M$_\odot$}

\newcommand{\tl}{$\langle\log t_\star\rangle_L$}

\title[Bimodality of the galaxy population]
         {Semi-empirical analysis of Sloan Digital Sky Survey galaxies: \\
         II. The bimodality of the galaxy population revisited}

\author[Mateus et al.]
{Ab{{\'\i}}lio Mateus$^{1}$\thanks{E-mail: abilio@astro.iag.usp.br},
Laerte Sodr\'e Jr.$^{1}$,
Roberto Cid Fernandes$^{2}$,
Gra\.zyna Stasi\'nska$^{3}$,
\newauthor
William Schoenell$^{2}$,
Jean M. Gomes$^{2}$\\
$^{1}$Departamento de Astronomia, IAG-USP, Rua do Mat\~ao 1226,
05508-090, 
S\~ao Paulo, Brazil\\
$^{2}$Depto.\ de F\'{\i}sica - CFM - Universidade Federal de 
Santa Catarina, Florian\'opolis, SC, Brazil\\
$^{3}$LUTH, Observatoire de Meudon, 92195 Meudon Cedex, France}

\begin{document}

\pagerange{\pageref{firstpage}--\pageref{lastpage}} \pubyear{2005}

\maketitle

\label{firstpage}

\begin{abstract}
We revisit the bimodal distribution of the galaxy population commonly
seen in the local universe. Here we address the bimodality observed in
galaxy properties in terms of spectral synthesis products, such as mean
stellar ages and stellar masses, derived from the application of this
powerful method to a volume-limited sample, with magnitude limit cutoff
$M(r) = -20.5$, containing about 50 thousand luminous
galaxies from the SDSS Data Release 2. In addition, galaxies are
classified according to their emission line properties in three distinct
spectral classes: star-forming galaxies, with young stellar
populations; passive galaxies, dominated by old stellar populations;
and, hosts of active nuclei, which comprise a mix of young and old
stellar populations. We show that the extremes of the distribution of
some galaxy properties, essentially galaxy colours, 4000 \AA\ break index,
and mean stellar ages, are associated to star-forming galaxies at one
side, and passive galaxies at another. We find that the mean
light-weighted stellar age of galaxies is the direct responsible for the
bimodality seen in the galaxy population. The stellar mass, in this view, has an additional role since most of the star-forming galaxies present in the local universe are low-mass galaxies. Our results also give support to the existence of a `downsizing' in galaxy formation, where massive galaxies seen nowadays have stellar populations formed at early times.
\end{abstract}



\section{Introduction}

It is usual to group galaxies in distinct types or classes, trying to 
find their general properties and to understand their 
differential formation and evolution. In the last few years, 
this galaxy `taxonomy' has been interestingly simplified.
As a by-product of recent  redshift surveys, the study of galaxy 
populations has quantitatively revealed the existence of a bimodal 
distribution in some fundamental galaxy properties, found both in 
photometric and spectroscopic data. This bimodality of galaxy
populations, despite its apparent simplicity, figures out as an
important key in our whole comprehension of the processes which drive
galaxy evolution.

Perhaps the most representative bimodal distribution seen in galaxy 
properties is that found in galaxy colours. Since photometric 
parameters are easily measured, their bimodal behaviour has been 
studied  for galaxies in the local universe by using both the 
Sloan Digital Sky Survey (SDSS) and the Two Degree Field Galaxy 
Redshift Survey (2dFGRS) data \citep{strateva01,hogg02,blanton03a,wild05}, 
and, for distant galaxies to $z \sim 1$, in the  COMBO-17 photometric 
redshift survey \citep{bell04} and in the DEEP galaxy redshift survey 
\citep{weiner05}. This behaviour also persists for even higher redshifts 
\citep*[$z \sim 1.4$][]{wiegert04}. As galaxy colours reflect the
star formation history of galaxies, a bimodality in their
distributions suggests that they have evolved through two different major
paths.

Amongst local galaxies, \citet{kauffmannII} have analysed the star 
formation history and its dependence on stellar mass for a large 
sample of SDSS galaxies. They have found a transition in the physical 
properties of galaxies at a stellar mass 
$M_\star \sim 3\times10^{10}$ M$_\odot$, which results in a bimodality 
between low-mass galaxies with low concentration indices typical of
discs and young stellar populations, and massive ones, with high 
concentration indices and old stellar populations. The connection 
between this bimodal distribution and that observed in
galaxy colours has been done, at least partially, by \citet{baldry04} 
in a work on the colour-magnitude diagram for a sample of SDSS galaxies. 
These authors have demonstrated that the transition in galaxy
properties occurs around $(1.5-2.2) \times 10^{10}$~M$_\odot$ for the red sequence 
and around $(2-3) \times 10^{10}$~M$_\odot$ for the blue sequence, close to the value
of $M_\star$ found by \citet{kauffmannII} using spectroscopic data.
This transition stellar mass is also associated to a shift in galaxy 
gas mass fractions, as pointed out by \citet{kannappan04}. Another important
characteristic of bimodality in galaxy colours is its dependence on
environment. This issue was addressed by \citet{balogh04} who have
complemented the analysis by \citet{baldry04} to account for, besides
luminosity dependences, the environmental dependence of the red and blue
galaxy fractions.

A bimodal behaviour is also seen in the star formation properties
of galaxies. From the 2dFGRS data, \citet{madgwick02} have derived 
a spectral parameter $\eta$ via a principal component analysis method 
that is strongly correlated 
with the star formation rate (SFR) of a galaxy. A bimodal distribution 
is clearly seen in this parameter \citep{wild05}, with a characteristic 
value that splits galaxies in two classes according to their star 
formation activity. \citet{brinchmann04} also find a similar 
distribution, showing that galaxies are from two kinds: 
one composed by concentrated galaxies, with low specific SFRs, and the other
composed by less concentrated ones with high specific SFRs.

In the first paper of this series on semi-empirical analysis of
galaxies \citep[][hereafter SEAGal I]{cid05a}
we have presented a spectral synthesis method able to derive the main 
physical properties of galaxies from their spectral data only. The 
reliability of this approach was investigated by three different ways:
simulations, comparison with other works, and an empirical analysis of 
the consistency of results for a SDSS galaxy sample. We find that 
spectral synthesis provides reliable physical parameters, mainly 
stellar ages and stellar metallicities. Stellar masses, velocity 
dispersions and extinction are also obtained through this method.
The analysis of residual spectra (observed minus synthesised) also
provides useful information on the overall properties of emission
line regions, like the nebular extinction and metallicities.

The joint spectroscopic and photometric data provided by SDSS,
in addition to the arsenal of physical parameters obtained by 
spectral synthesis, turns out as a great opportunity to study galaxy
populations by analysing both spectral and structural classifications. 
In this work, we will produce a spectral classification,
initially based on the emission line properties of galaxies and
the presence of nuclear activity, and subsequently we will 
compare this classification with a structural (morphological) one, 
which uses SDSS photometric information to separate galaxies 
between early and late-types. The bimodality of the galaxy
populations will be analysed here with the starting point placed in 
these classifications.

This paper is organised as follows. Section \ref{sec:Synthesis}
presents an overview of our synthesis method, detailed in SEAGal I, 
focusing on the main outputs resulting from its application to a 
volume limited sample of SDSS galaxies. Section \ref{sec:Classification}
describes the classification scheme used along this work, based on a 
diagnostic diagram of emission line ratios. Section \ref{sec:Bimodality}
discusses the bimodal character of the galaxy population, with emphasis 
in the emission line properties of galaxies. In Section 
\ref{sec:Discussion} we discuss the main findings of this 
work. Finally, Section \ref{sec:Conclusions} summarises our results.


\section{The spectral synthesis method}\label{sec:Synthesis}

In this section we present an overview of the spectral synthesis
method used in this work and its main output. We also present the
volume limited sample of SDSS galaxies analysed here, as well as our
procedure for measuring emission line intensities from the residual
spectra.

\subsection{The method}\label{sec:Synthesis_method}

In this work we use the STARLIGHT code to derive stellar population properties
directly from the galaxy spectra (continuum and absorption lines
only). STARLIGHT is built upon computational techniques originally developed
for empirical population synthesis with additional ingredients from
evolutionary synthesis models. In brief, we fit an observed spectrum with a
combination of $N_\star$ Simple Stellar Populations (SSP) from the
evolutionary synthesis models of \citet{bc03} (hereafter BC03).  We use the
``Padova 1994'' tracks recommended by BC03 and a Chabrier (2003) IMF between
0.1 and 100 M$_\odot$. (Experiments with a Salpeter IMF with the same mass
limits were also performed, yielding nearly identical results except for
stellar masses, which are 1.5 times larger than for a Chabrier IMF.)
Extinction is modelled as due to foreground dust, with the reddening law of
\citet*{cardelli89} with $R_V=3.1$, and parameterised by the V-band extinction
$A_V$. Line of sight stellar motions are accounted for using a Gaussian
distribution.  The fits are carried out with a simulated annealing plus
Metropolis scheme, with regions around emission lines and bad pixels excluded
from the analysis (see SEAGal I for details).

Since SEAGal I, STARLIGHT has undergone a major revision. The main
technical change is that whereas in SEAGal I we worked with a single
Markov chain all the way from an initial guess to the final model, we
now run several parallel chains. At each level of the annealing scheme
(i.e., for each ``temperature'') we run the chains until they satisfy a
convergence criterion similar to that proposed by \citet{gelman92}.
Adaptive step sizes are employed to improve
efficiency. These and other details will be described in
a future communication accompanying a public version of STARLIGHT.

A second and more important difference with respect to SEAGal I is
that we now use a larger base of SSPs, comprising $N_\star = 150$
elements of 25 different ages between 1 Myr and 18 Gyr, and 6
metallicities: $Z=0.005, 0.02, 0.2, 0.4, 1$ and 2.5 $Z_\odot$. The age
grid was chosen by exploring an implicit assumption in spectral
synthesis with a discrete base, namely, that the spectrum of a
population with age $t$ between $t_j$ and $t_{j+1}$ is well
represented by a linear combination of the $j$ and $j+1$ SSPs. We
started from the set of 15 ages in the SEAGal I base, and then fitted
each of 221 BC03 SSP spectra for the same $Z$ with the two components
which bracket its age. This allows us to identify SSPs whose spectra
are poorly interpolated by combinations of the closest elements in the
base, and thus should be included in a finer base. Following this
procedure (graphically illustrated in \citealt{leao06}), we arrived at a
set of 25 ages. The difference between original and interpolated SSP
spectra with this base is better than 1 per cent on average, exceeding 5
per cent only for a few $t < 10^7$ yr models at $Z \le 0.2 Z_\odot$,
corresponding to short-lived phases when massive red-supergiants show up.

\begin{figure}
\centerline{\includegraphics[scale=0.45]{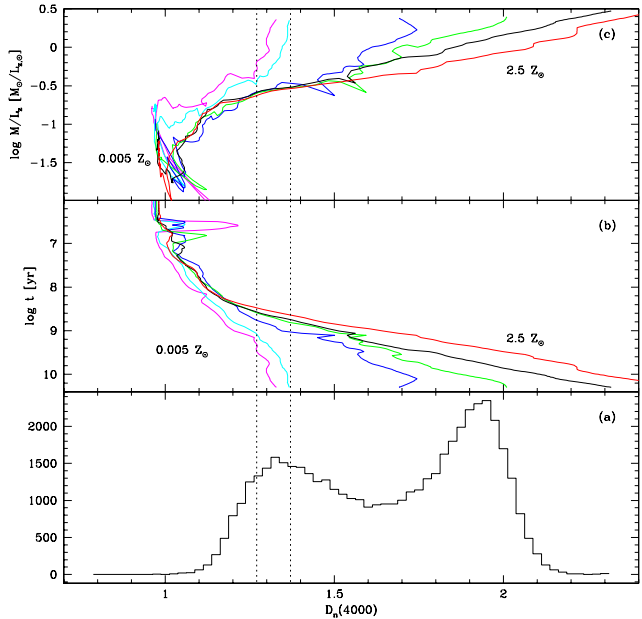}}
\caption{(a) Observed distribution of the 4000 \AA\ break index
for our volume limited sample.  (b) Evolution of D$_n$(4000) for SSPs
of 6 different metallicities. From right to left: $Z = 2.5$, 1, 0.4,
0.2, 0.02 and 0.005 $Z_\odot$. (c) Relation between their D$_n$(4000) and $z$-band mass-to-light ratio for SSPs of 6 different metallicities. From right to
left: $Z = 2.5$, 1, 0.4, 0.2, 0.02 and 0.005 $Z_\odot$. The dotted lines mark
the location of the lower peak in the observed D$_n$(4000) distribution.}
\label{fig_Dn4000_x_SSPs}
\end{figure}

The refinement in the age grid is not as relevant as the inclusion of
lower metallicity components in the base. Whereas the $N_\star = 45$
base in SEAGal I comprised only $Z \ge 0.2 Z_\odot$, we now allow for
$Z=0.005$ and $0.02 Z_\odot$.  These very metal poor SSPs are bluer
than those of $Z \ge 0.2 Z_\odot$ SSPs, particularly at old ages. Such
low $Z$ populations, if present in a galaxy but ignored in the fits,
may lead to an underestimation of ages, as the code tries to compensate
for their lack with younger components of higher $Z$ due to the
age-$Z$ degeneracy.  To illustrate this point, Fig
\ref{fig_Dn4000_x_SSPs} shows the predicted evolution of the 4000 \AA\
break index $D_n(4000)$ (Section \ref{sec:spectral_vs_structural}) and its
observed distribution for the sample described below.  The peak at
$D_n(4000) \sim 1.3$ matches well the value reached by
$Z \le 0.02 Z_\odot$ SSPs of ages from 1 Gyr to over 10 Gyr. To reproduce
this value with $Z \ge 0.2 Z_\odot$ one must either accept that these
galaxies formed less than $\sim 1$ Gyr ago or else invoke more continuous
star-formation regimes (exponentially decaying, multiple bursts or other
variants). Whereas the latter is in fact a likely explanation for
galaxies with low $D_n(4000)$, our point here is that by excluding low
$Z$ components from the base one would be in practise {\it imposing}
this interpretation {\it a priori}.  Besides this ``mathematical''
argument, the mere fact that very metal poor stars exist provides
physical motivation to include them in the base.  Having said that,
one must also recall that the synthetic SEDs of such
populations are affected by a number of caveats, as discussed by BC03.

One must also realise that the inclusion of 0.02 and 0.005 $Z_\odot$
SSPs is bound to have an effect on the stellar masses. This is
illustrated in the top panel of Fig \ref{fig_Dn4000_x_SSPs}, which
shows that the mass-to-light ratio in the $z$-band of $\sim 10$ Gyr,
0.02 and 0.005 $Z_\odot$ SSPs is up to one order of magnitude larger
than for $Z \ge 0.2 Z_\odot$ models of similar $D_n(4000)$. To monitor
the magnitude of this difference we have also fit the data with a base
excluding the 0.02 and 0.005 $Z_\odot$ SSPs. Throughout the paper we
will mention results obtained with this alternative base whenever they
differ substantially from the ones obtained with the full set of
metallicities. We anticipate that such changes are mostly
quantitative; they do not affect the overall qualitative conclusions of
our analysis.

Another difference with respect to the base in SEAGal I is that we now
allow for SSPs as old as 18 Gyr, which is formally inconsistent with
the $\sim$ 13.5 Gyr old Universe implied by our adopted cosmology, $H_0=70$
km s$^{-1}$ Mpc$^{-1}$, $\Omega_M=0.3$ and $\Omega_\Lambda=0.7$.  We
relax the cosmological limit merely to avoid the crowding of mean
stellar ages at the upper age limit (visible, for instance, in SEAGal
I's Fig.\ 13b). Given the uncertainties in stellar evolution,
cosmology, and the spectrophotometry measurements,
we do not regard this inconsistency as critical, and, in
any case, we stress that this choice has no consequence for the main
conclusion of this paper.

The result of this procedure is a list of parameters for each galaxy.
As explained in SEAGal I, instead of working with the full population
vector, whose individual components are hopelessly plagued by
mathematical and astrophysical degeneracies, we describe the results
of the spectral synthesis in terms of a small number of robust
parameters. The most important for the discussion in this paper are:
a) the dust-corrected present stellar mass inside the fibre; the total stellar
mass ($M_\star$) is computed \textit{a posteriori} after correcting
for the fraction of luminosity outside the fibre assuming that the
galaxy mass-to-light ratio does not depend on the radius (but see
Appendix A); b) the mean stellar age, weighted either by light,
$\langle\log t_\star\rangle_L$, or by mass $\langle\log
t_\star\rangle_M$; c) the mean stellar metallicity, light and
mass-weighted, $\log \langle Z_\star\rangle_L$ and $\log \langle
Z_\star\rangle_M$, respectively; and d) the V-band stellar extinction
$A_V^\star$. A summary of the uncertainties in the main parameters that will be used
in this work is shown in
Table~\ref{tab:Parameter_Uncertainties}. These uncertainties were
determined in SEAGal I through the difference between output and input
values of the corresponding quantity, as obtained from simulations
with different signal-to-noise ratios.  In addition to this list of
parameters, by subtracting the synthesised spectrum from the observed
one, we get a ``pure-emission'' residual spectrum, which is useful for
analysis of the galaxy emission lines. These are discussed next.

\begin{table}
\centering
\begin{tabular}{lccccc}
\hline
   Parameter     &
\multicolumn{5}{c}{$S/N$ at $\lambda = 4020$ \AA} \\
    &
  5 &
10 &
15 &
20 &
30 \\ \hline
$\log M_\star$ 	                &  0.11 &  0.08 &  0.06 &  0.05 &  0.04 \\
$\langle\log t_\star\rangle_L$ 	&  0.14 &  0.08 &  0.06 &  0.05 &  0.04 \\
$\langle\log t_\star\rangle_M$ 	&  0.20 &  0.14 &  0.11 &  0.10 &  0.08 \\
$\log \langle Z_\star\rangle_L$	&  0.15 &  0.09 &  0.08 &  0.06 &  0.05 \\
$\log \langle Z_\star\rangle_M$ &  0.18 &  0.13 &  0.11 &  0.09 &  0.08 \\
$A^\star_V$          		&  0.09 &  0.05 &  0.03 &  0.03 &  0.02 \\
\hline
\end{tabular}
\caption{Summary of parameter uncertainties. Each row corresponds to a parameter constructed by combining the parameters given by the synthesis. The different columns list the rms difference between output and input values of the corresponding quantity, as obtained from simulations with different signal-to-noise ratios (see SEAGal I for details). The units are dex for logarithmic quantities and mag for $A^\star_V$.}
\label{tab:Parameter_Uncertainties}
\end{table}

\subsection{Application: SDSS data}

We apply our synthesis method to a large sample of SDSS galaxies
aiming to investigate the properties and the origin of the bimodal
galaxy distribution. The SDSS spectra cover a wavelength range of
3800--9200 \AA, have a mean spectral resolution $\lambda/\Delta\lambda
\sim 1800$, and were taken with 3 arcsec diameter fibres.  The
spectroscopic sample studied here is the same discussed in SEAGal
I. Briefly, we selected a volume limited sample from the SDSS Data
Release 2 \citep[DR2][]{DR2}, with a redshift range of $0.05 < z <
0.1$, corresponding to an absolute magnitude limit cutoff $M(r) = -20.5$, or
$M^*(r) + 1$. The absolute magnitudes adopted here are $k$-corrected
\citep{blanton03kcorrect}.  We also restricted our sample to objects
for which the observed spectra show a signal-to-noise ratio greater
than 5 in the $g$, $r$, and $i$ bands. In addition, we have detected a
minor fraction of galaxies with multiple spectral data, from which we
have excluded those with smaller signal-to-noise ratios. All these
criteria leave us with a sample containing 49917 galaxies,
corresponding to a completeness level of 98.5 per cent.

The spectra are corrected for Galactic extinction with the maps of by
\citet*{schlegel98} and the extinction law of \citet{cardelli89},
shifted to the rest-frame and resampled from 3400 to 8900 \AA\ in steps
of 1 \AA. Notice that this is a slightly wider range than that
employed in SEAGal I (3650--8000 \AA), although not all galaxies have
data close the edges of this new spectral range.

\subsubsection{Masks}

A further novelty is that now each galaxy has its own emission line
mask, as opposed to the single general mask for all galaxies used in
SEAGal I. General masks may unduly throw away data when emission
lines do not exist, and, conversely, ignore lines which are normally
weak but may occasionally be strong enough to deserve being masked.
(e.g. \heii$\lambda$4686).  The individual masks are constructed from an
initial STARLIGHT fit with a general mask. The strongest lines are
fitted in the residual spectrum (observed minus model) using the code
described in the next section, and the information on line widths and
velocity off-sets are used to search for other emission lines out of a
list of 52 lines in the 3400 to 8900 \AA\ range. When present, these
lines are masked in windows whose sizes are defined by the $\lambda$'s
at which their assumed Gaussian profiles become weaker than 1/2 the
local rms noise flux density. Visual inspection of hundreds of cases
shows that this scheme is able to identify and mask even weak emission
lines satisfactorily.

Besides emission line masks, we also exclude four other windows from
the fits: 5880--5906 \AA, to skip the Na D $\lambda\lambda$5890,5896
doublet, which is partly produced in the interstellar medium; the
6850--6950 and 7550--7725 \AA, for which BC03 had to resort to
theoretical spectra due to problems in these ranges in the STELIB
library. Curiously, when residual spectra are averaged over thousands
of galaxies, these two windows stand out as strong ``emission''
features which are clearly spurious \citep{gomes05}.  The last window, in
the 7165--7210 \AA\ range, shows a similar systematic broad residual
in emission, so we have masked it out too, even though it is not
listed as problematic by BC03.

\subsection{Emission line measurements}\label{sec:line_measurements}

One of the products of our spectral synthesis procedure is a residual
spectrum suitable for measurements of emission lines, as the stellar
continuum and absorption features are well modelled. Even weak lines,
as the \Oiiit~auroral line, can be easily measured (when present, of
course). Indeed, such an approach has been adopted by other authors
\citep[e.g.][]{tremonti04} to investigate SDSS emission line spectra.

We have developed a code to measure the intensity of main emission
lines from the residual spectrum by fitting Gaussian functions to the
line profiles, characterised by three parameters: width, offset (with
respect to the rest-frame central wavelength), and flux. Lines from
the same ion are assumed to have the same width and
offset. Additionally, \oiii$\lambda5007$/\oiii$\lambda4959 = 2.97$ and
\nii$\lambda6584$/\nii$\lambda6548 = 3$ flux ratio constraints are
imposed.  Furthermore, only lines detected with signal-to-noise ratio
($S/N$) greater than 3 are considered in the analysis. This is the
same value adopted by \citet{brinchmann04} in their study of physical
properties of star-forming SDSS galaxies, and is 1$\sigma$ greater
than the detection limit adopted by \citet{miller03} in an
environmental analysis of active galactic nuclei, also in the
SDSS. The lines that are currently being measured by our code include
\oii$\lambda\lambda$3726,3729, H$\delta$, H$\gamma$,
\oiii$\lambda$4363, H$\beta$, \oiii$\lambda\lambda$4959,5007,
\oi$\lambda$6300, \nii$\lambda$6548, H$\alpha$, \nii$\lambda$6584 and
\sii$\lambda\lambda$6716,6731.  For each emission line, our code
returns the rest-frame flux and its associated equivalent width (EW),
the line width, the velocity displacement relative to the rest-frame
wavelength, and the $S/N$ of the fit. In the case of Balmer lines, the
underlying stellar absorption is also measured directly from the
synthetic spectra, yielding absorption fluxes and equivalent widths.

Additionally, we have also identified broad emission lines in the
residual spectra, characteristic of galaxies hosting Seyfert 1 nuclei,
following a procedure similar to that of \citet{veron01}. As a result
of this analysis we have identified 335 objects with significant broad
components.


\begin{figure}
\centerline{\includegraphics[scale=1.0]{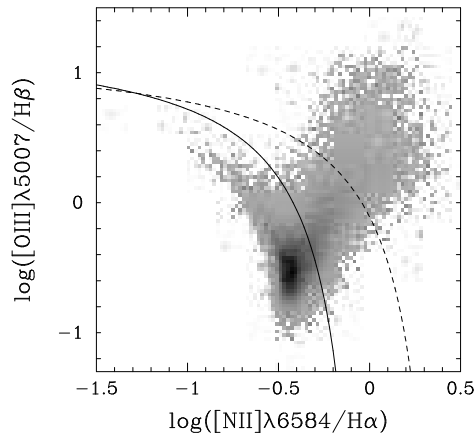}}
\caption{The BPT diagram for our subsample of galaxies with emission
lines measured with $S/N$ greater than 3. The solid line is the curve
defined by \citet{kauffmannAGN} and the dashed one is the curve defined by
\citet{kewley01} (see text for details). In this and other figures, the linear grey scale level represents the number of galaxies in each pixel, darker pixels having more galaxies.}
\label{fig_bpt}
\end{figure}

\section{Definition of spectral classes of galaxies} \label{sec:Classification}

In this section we present the galaxy types that will be used
hereafter. This galaxy classification is based 
on spectral properties, i.e. on the presence or absence of emission lines
and on their ratios in the case they are present in the spectra.

\subsection{Galaxy classification}

In general, the classification of narrow emission line galaxies 
is made according to the mechanism by which lines are produced,
grouping galaxies into two types. The first group is composed by 
normal star-forming galaxies, with \hii region-like spectra, in which 
the gas is photoionised by young,
hot OB stars. The second one is formed by galaxies hosting active galactic nuclei
(AGN), which produce a much harder radiation field ionising the surrounding gas.

These emission line galaxies are generally classified with the help of 
diagnostic diagrams formed by line ratios of the strongest lines present in
their spectra. Here we will use the \Oiii/H$\beta$ versus
\Nii/H$\alpha$ diagram firstly proposed by \citet{bpt}  and updated 
by \citet{veilleux87}; hereafter we will refer to it as the BPT diagram. 
In it, galaxies form two distinct branches, which look like the wings of 
a seagull. The left wing contains star-forming galaxies and the observed 
sequence corresponds to a change in metallicity of the \hii regions 
emitting the lines. The right wing appeared clearly only
with the most recent galaxy surveys \citep[e.g.][]{kauffmannAGN} 
and corresponds to galaxies hosting active nuclei; photoionization models 
show that galaxies in this wing cannot be ionised only by radiation from
massive stars, an additional heating/ionising source is necessary to
explain the observed line ratios \citep[e.g.][]{kewley01}.

\begin{figure}
\resizebox{\columnwidth}{!}{\includegraphics{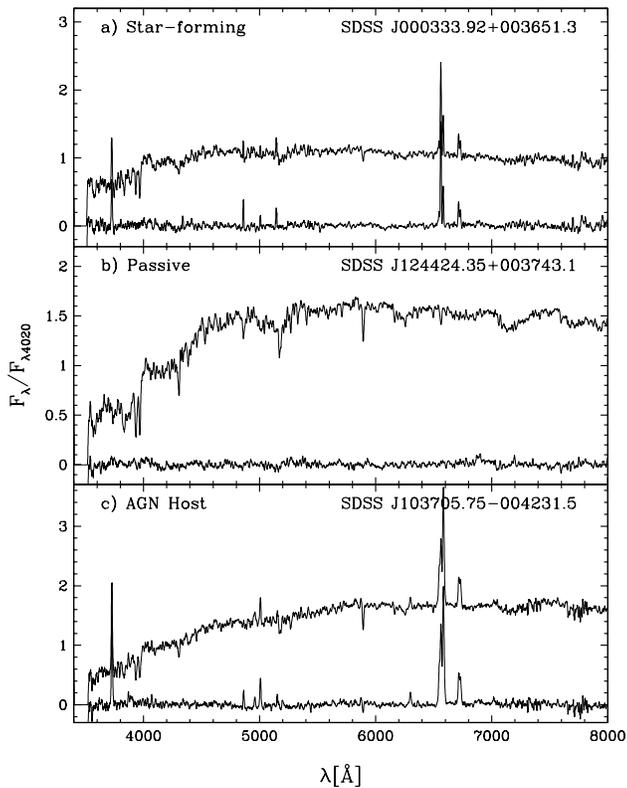}}
\caption{Examples of galaxy spectra of the spectral classes analysed 
in this work. The residual spectra are also shown as thin lines at the
bottom of each panel. These spectra have been smoothed in order to improve appearance.}
\label{fig_selected_spectra}
\end{figure}

The BPT diagram for our sample is shown in Fig.~\ref{fig_bpt}.
We have considered only those galaxies having  a signal-to-noise
ratio in all four emission lines greater than 3; 
this limit results in a subsample containing $23080$ narrow emission
line galaxies. Here we have not considered galaxies with spectra 
containing broad emission lines (following Section \ref{sec:line_measurements} above).
After an analysis of SDSS data, \citet{kauffmannAGN} have defined
a curve (shown in Fig.~\ref{fig_bpt} as a solid line) in the BPT
diagram to distinguish pure star-forming galaxies from
those contaminated by nuclear non thermal emission. Although the 
exact position of this curve could be questionned \citep[see][]{brinchmann04, stasinska06},
we have adopted it for easy reference to recently published work \citep[e.g.][]{hao05,brinchmann04}
to distinguish normal star-forming galaxies (below the curve) from
galaxies hosting an active nucleus (above it). In Fig.~\ref{fig_bpt} 
we also show the curve given by \citet{kewley01}, which we use in
Sect. 4.2.1 to define galaxies with `composite' spectra, as done by 
\citet{brinchmann04}.

The normal star-forming galaxies in our sample show significant amount of
star formation (more than 99 per cent have EW(H$\alpha$)~$>5$\AA).
Note that as we are studying a volume limited sample built with an
absolute magnitude cutoff $M(r) < -20.5$, in this diagram we do not 
see the faint star-forming galaxy population that inhabits the extremes 
of its left wing (corresponding to a line ratio of 
\nii/H$\alpha$ $\la 0.1$).

Galaxies hosting active nuclei (hereafter AGN hosts), on the other
hand, populate the upper right wing of the BPT diagram. 
In addition, we do not distinguish these galaxies from the `composite'
or `transition' galaxies, which have a substantial contribution of 
their \Ha emission due to active nuclei, and populate the region near 
the solid curve shown in Fig.~\ref{fig_bpt}.

The objects that do not appear in  Fig.~\ref{fig_bpt} because 
of insufficient S/N in at least one of the four relevant lines also receive
a classification. For instance, galaxies that host active nuclei are 
also identified if they have only the emission lines of H$\alpha$ and 
\nii\ measured with 3$\sigma$ confidence \citep[e.g.][]{coziol98,miller03}.
We have selected these galaxies by considering the limit 
log~(\nii/H$\alpha) > -0.2$, and classified them as AGN hosts. We note that the fraction of galaxies with active nuclei identified by this way is about 54 per cent of the total of AGNs hosts in our sample. Additionally, galaxies with broad emission lines which host Seyfert 1 nuclei, are naturally added to the spectral class of AGN hosts.
Galaxies without evidence of significant star formation,
whose spectra do not show the emission lines of H$\alpha$ and H$\beta$,
or show them with equivalent widths smaller than 1 \AA, are classified
as passive galaxies. Note that with this definition we are not including the
fraction of elliptical galaxies which show evidence of star formation 
activity \citep[e.g.][]{nakamura04,fukugita04} in our subsample of 
passive galaxies. The remaining galaxies, for which we can not give a classification
based on the BPT diagram, tend to include mainly star-forming and 
passive galaxies. These unclassified galaxies represent a
reduced fraction of galaxies in our sample (about 3 per cent) and
will not be analysed in this work.

A summary of the spectral classes described above, which will be 
analysed in the following sections, is shown in Table \ref{table_summary}.
The number and per percentage of each spectral class, as well as the median
values of their concentration indices, colours, 4000 \AA\ break, mean
light-weighted stellar ages and stellar masses, are shown in this table.
In Fig.~\ref{fig_selected_spectra} we show some examples of galaxy 
spectra for the distinct classes of galaxies discussed above.
In each panel is also shown the residual spectrum after removing
the model obtained by our synthesis method. These spectra correspond 
to galaxies with both stellar age and stellar mass equal to their
median values for each class, therefore they are very illustrative of 
the different spectral features which characterise the classes we
discuss here.

\begin{table*}
\begin{tabular}{cccccccc}
  \hline
  Class     & Number & Per cent & $C$ & $(u-i)$ & $D_n(4000)$ & \tl  & $\log
M_\star/M_\odot$ \\
  \hline
All         & 49917  & 100.00 & 2.67 & 2.76 & 1.69 & 9.53 & 10.79\\
Star-forming& 16108  &  32.27 & 2.25 & 2.10 & 1.35 & 8.91 & 10.56\\
Passive     & 10485  &  21.00 & 2.97 & 3.06 & 1.93 & 9.86 & 10.87\\
AGN Hosts   & 21733  &  43.54 & 2.77 & 2.88 & 1.78 & 9.66 & 10.90\\
Unclass.    & 1591   &   3.19 & 2.67 & 2.83 & 1.73 & 9.58 & 10.77\\
\hline
\end{tabular}
\caption{Summary of the spectral classes defined in this work and the
median values of their mean light-weighted stellar ages and stellar
masses.}
\label{table_summary}
\end{table*}

\subsection{Sources of bias}

\begin{table}
\begin{tabular}{ccccccc}
  \hline
  $S/N$ & Star-forming & Passive & AGN Hosts & Unclass. \\
  \hline
2& 16024 \, 32.1&  8626 \, 17.3& 24321 \, 48.7&  946 \quad 1.9\\
3& 16108 \, 32.3& 10485 \, 21.0& 21733 \, 43.5& 1591 \,    3.2\\
5& 17034 \, 34.1& 12998 \, 26.0& 17382 \, 34.8& 2503 \,    5.0\\
  \hline
\end{tabular}
\caption{Number and percentage of objects in each spectral class as a
function of the limit in $S/N$ when considering a line in emission. Here
we are using a $S/N$ limit equals to 3.}
\label{table_classes}
\end{table}

\begin{table}
\begin{tabular}{ccccc}
\hline
  $S/N$ limits & Star-forming & Passive & AGN Hosts & Unclass. \\
\hline
  2 and 3 & 99.99 & 99.99 & 50.97 & 31.38 \\
  3 and 5 & 99.92 & 99.99 &  0.30 & 25.12 \\
  \hline
\end{tabular}
\caption{Kolmogorov-Smirnov (KS) probability parameter for
redshift distributions when comparing the $S/N$ limits 
of 2, 3 and 5 for considering a line in emission. Results are shown 
for each spectral class. Low values of the KS probability imply 
statistically different distributions.}
\label{table_classes_KS}
\end{table}

\begin{figure}
\centerline{\includegraphics[scale=0.35]{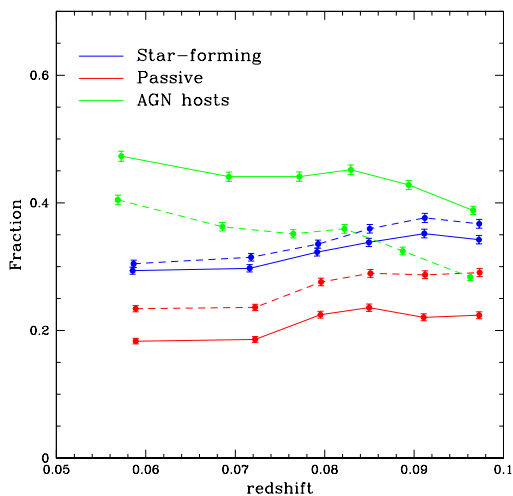}}
\caption{Testing aperture effects. The fraction of galaxies
in each spectral class is shown in different bins of redshift. Solid lines are for fractions obtained with the $S/N > 3$ limit for detection of emission lines, whereas dashed lines are for the limit $S/N > 5$. The error bars follow a Poisson statistics.}
\label{fig_aperture_effects}
\end{figure}

As mentioned in Section \ref{sec:line_measurements}, in this work
we are imposing a limit in the $S/N$ of the measured lines 
in order to consider them in emission or not. The choice
of this limit affects the distribution of objects in each spectral 
class, as we shown in Table \ref{table_classes}, where the number and 
percentage of galaxies in our classes are listed for three limiting
values of the $S/N$. For a change in $S/N$ going from 2 to 5, 
the proportion of star-forming galaxies increases by only 6 per cent,
but that of passive galaxies increases by 50 per cent. In contrast,
the proportion of active nuclei hosts decreases by about 28 per cent.
It is interesting to note that the fraction of unclassified galaxies is
always smaller than 5 per cent.

We have investigated how the choice of the $S/N$ limit affects 
our analysis by looking for differences in the redshift distributions 
of our classes. This approach can also test possible aperture bias in 
our class definition. Since SDSS spectra are taken with only 3 arcsec 
fibre diameter, for the nearest galaxies they might tend to sample 
mainly the bulge component, increasing the fraction of non emission 
line galaxies (especially passive ones). Therefore, the redshift 
distributions for each spectral class defined by using different 
$S/N$ limits, will be similar if the aperture bias does not play a 
significant role in our sample, or it does in an uniform way 
independently of the $S/N$ value. In Table \ref{table_classes_KS} we 
show the Kolmogorov-Smirnov (KS) probability parameter for the redshift 
distributions of our distinct spectral classes. We have compared the 
distributions of classes defined by using a $S/N$ limit equals to 2 
and 5, with those of our classes defined by the adopted value of 
$S/N = 3$. The values of the KS probability are very large for passive
and star-forming galaxies, indicating statistically similar
distributions for these spectral classes. In the case of AGN hosts, the
value of the KS probability parameter is large when comparing $S/N$
limits of 2 and 3, but it is very small for $S/N$ limits of 3 and 5,
indicating statistically different redshift distributions when
comparing these $S/N$ limits. For unclassified galaxies the KS
probability values are larger than 25 per cent in the two cases
investigated here.

In Fig.~\ref{fig_aperture_effects} we show the redshift dependence of the fraction of galaxies in each spectral class obtained by considering two values of the $S/N$ limit to detect emission lines (3 and 5). The expected effects due to aperture bias would tend to, for
instance, increase the fraction of star-forming galaxies relative to
that of passive along the redshift range of our study. In fact, what we
note in this figure is an increment in the fraction of star-forming
galaxies with increasing redshift. In addition, for lower redshifts, we
also observe a higher fraction of AGN hosts 
probably as a result of the significant sampling of the central regions
of nearby galaxies. In contrast, the fraction of passive galaxies does
not significantly increase in the regime of low redshifts. Following
these trends, we infer that our classes seem to be affected by aperture
bias in the sense that the fraction of star-forming galaxies increases
with redshift, and that of AGN hosts decreases with $z$. Some aspects
of this effect in our sample are discussed in Appendix~\ref{appendix:aperture_bias}.


\begin{figure*}
\resizebox{\textwidth}{!}{\includegraphics{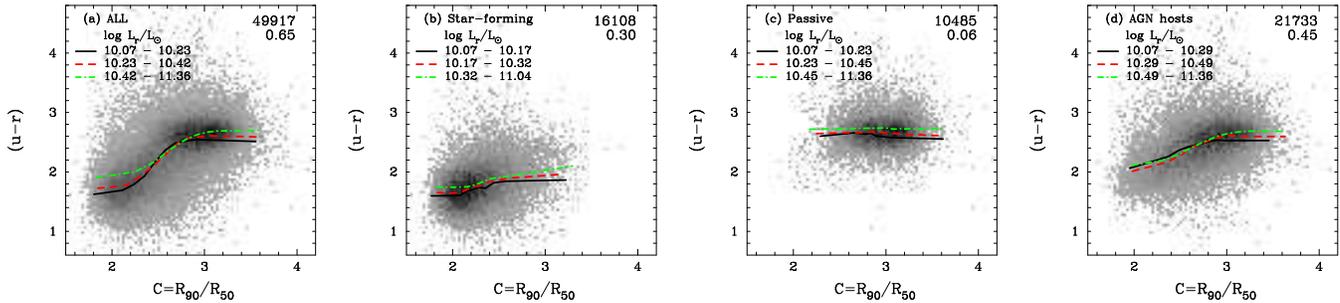}}
\caption{Concentration index versus $(u-r)$ colour for all galaxies 
in our sample (panel a) and for our distinct spectral classes 
(panels b-d). In this and in other figures of this paper, the top
numbers on the right are the number of galaxies in each panel and
the Spearman rank correlation coefficient ($r_S$). The median relation
between colour and concentration index in three bins of galaxy 
$r$-band luminosity containing the same number of galaxies is shown as
distinct lines. The ranges of each bin are shown in
the top-left legend.}
\label{CIcolour}
\end{figure*}

\section{Bimodality of the galaxy population}\label{sec:Bimodality}

In this section, we discuss the characterisation of the two galaxy populations that inhabit the local universe. Our starting point is the definition of the spectral classes discussed in the last section.

\subsection{Spectral versus structural galaxy types}
\label{sec:spectral_vs_structural}

Many works on the SDSS data have found that the concentration
index ($C$), defined as the ratio of Petrosian $R_{90}$ to $R_{50}$ 
radii in $r$-band can be used to separate early and late-type galaxies \citep{shimasaku01,strateva01,goto02}, since early-type
galaxies tend to have light profiles more centrally concentrated than the
late-types \citep{morgan58}. \citet{shimasaku01} studied this
parameter for a sample of morphologically classified bright galaxies
from SDSS. They found a strong correlation between $C$ and
morphological types, suggesting that this parameter can be used to 
classify galaxies morphologically \citep*[see also][]{doi93,abraham94}. 
However, they also noted that it is difficult 
to construct a pure early-type galaxy sample based only on the 
concentration index, since the resulting sample has $\sim 20$ per cent  
contamination by late-type galaxies. \citet{strateva01} 
also studied the reliability of the concentration index to
separate early and late-type galaxies. They adopt a value of 
$C = 2.63$ as a morphological separator (different from  
Shimasaku's value of $C \sim 3$). On the other hand, since galaxy 
colours are a more conventional estimator of galaxy types 
\citep[e.g.,][]{sandage86}, Strateva et al. concentrate their analysis 
on the $(u-r)$ colour, finding that $(u-r) = 2.22$ clearly separates 
early (E, S0, and Sa) and late (Sb, Sc, and Irr) morphological types, 
with the $(u-r)$ colour also expected to correlate with Hubble type.

How are these parameters associated with the spectral classes we
defined before? In Fig.~\ref{CIcolour} we show the relation of 
these two main morphological identifiers, $C$ and $(u-r)$,
for all objects in our sample and discriminated according to
our distinct spectral classes (except for unclassified galaxies). 
In this figure, we also plot as lines with different shapes 
the median values of $(u-r)$ colour along bins of concentration index
for three luminosity ranges containing the same number of galaxies.
As found by other works \citep[e.g.][]{strateva01}, the concentration
index and $(u-r)$ colour present an evident correlation (quantified by the
Spearman non-parametric correlation coefficient, $r_S = 0.65$) when all
objects are considered. Star-forming galaxies also show a correlation
between these parameters, in a less significant level, in the sense that
red star-forming objects tend to be more concentrated than blue ones.
On the other hand, passive galaxies do not show a correlation; 
the colours of these objects are almost independent of their concentration, 
in all ranges of galaxy luminosity. Indeed, for passive galaxies in our
sample we have the following mean values and dispersions:
${\langle C \rangle}= 2.96 \pm 0.08$ and ${\langle u-r \rangle}=2.71 \pm 0.22$.
In Fig.~\ref{CIcolour} we also note that star-forming and passive 
galaxies are responsible for the distinction of the two main groups of galaxies 
in the colour--$C$ diagram. The class of AGN hosts is composed by mixtures 
of blue and red galaxies, with a large range in the concentration 
parameter.

\begin{figure*}
\resizebox{\textwidth}{!}{\includegraphics[scale=1]{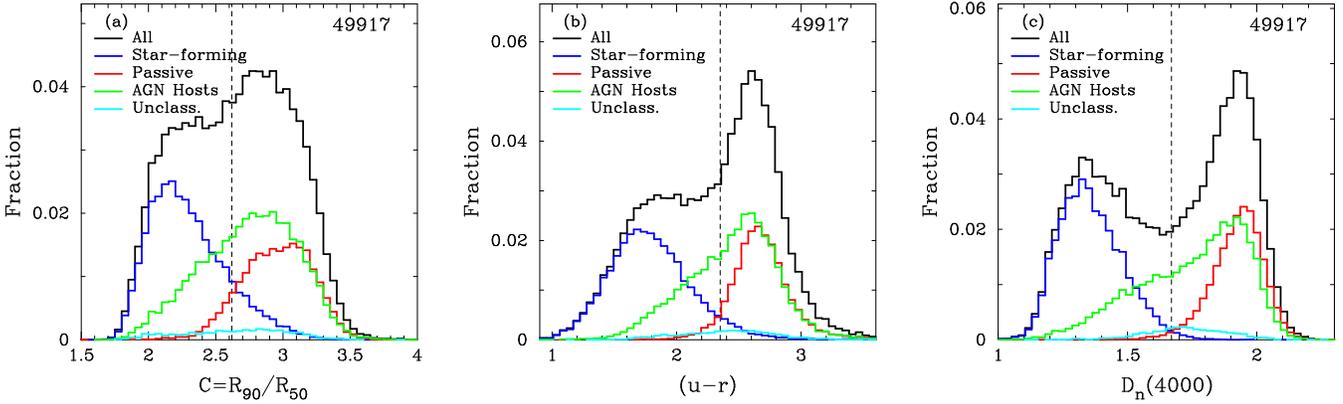}}
\caption{Distributions of (a) concentration index $C$, (b) $(u-r)$
colour, and (c) $D_n(4000)$ index, for all galaxy sample,
star-forming, passive and AGN hosts spectral classes; the distributions for
unclassified objects are also shown. The vertical dashed lines are the best
population separators when we consider only distributions of star-forming and passive galaxies.}
\label{fig_cumulative_colour_CI}
\end{figure*}

We show in Fig.~\ref{fig_cumulative_colour_CI}a-b  the 
distributions of concentration index and $(u-r)$ colour for our spectral 
classes (also including unclassified galaxies). In these figures, we can
easily see the dominance of star-forming galaxies in the blue and less 
concentrated galactic population; passive galaxies, contrarily comprise 
the reddest and more concentrated galaxies. We also note that AGN hosts
occupy an intermediate locus in both colour and concentration index
distributions, showing again the mix of populations present in this
spectral class.

From the distributions shown in Fig.~\ref{fig_cumulative_colour_CI},
we can obtain an optimal value that separates those two extreme classes
formed by star-forming and passive galaxies. Following 
\citet{strateva01}, we can define two parameters, reliability and 
completeness, which are used to do this task. The reliability 
($\mathcal R_{\rm SF}$ and $\mathcal R_{\rm P}$, for star-forming and 
passive spectral classes, respectively) is the fraction of
galaxies from a given spectral class that are correctly classified by
using the optimal value, while the completeness
($\mathcal C_{\rm SF}$ and $\mathcal C_{\rm P}$) is the
fraction of all galaxies of a given spectral class that are actually
selected. In this way, we can find the optimal value that maximise the
product $\mathcal{C_{\rm SF} R_{\rm SF} C_{\rm P} R_{\rm P}}$
\citep[c.f.][]{baldry04}. We find that $C=2.62$ and $(u-r) = 2.35$ are the
optimal separators among star-forming and passive galaxies (close to the
values obtained by Strateva et al. to classify galaxies into early and
late-types), with reliability and completeness parameters (in per cent) of
$\mathcal R_{\rm SF} = 86.0$, $\mathcal R_{\rm P} = 89.8$,
$\mathcal C_{\rm SF} = 93.0$ and $\mathcal C_{\rm P} = 80.3$ for
concentration index and
$\mathcal R_{\rm SF} = 93.6$, $\mathcal R_{\rm P} = 93.8$,
$\mathcal C_{\rm SF} = 96.0$ and $\mathcal C_{\rm P} = 90.4$ for
$(u-r)$ colour.
Here it is important to stress that we are not accounting for 
the variation of the colour divider along the magnitude range of our 
sample. Indeed, as shown by \citet{baldry04} the colour value which 
divides the blue and red populations should slightly decrease with decreasing
galaxy luminosity.

Another parameter useful to distinguish between early and late-type
galaxies is the 4000 \AA\ break, which is small for galaxies
with younger stellar populations, and large for older galaxies 
(as shown in Fig. 13 of SEAGal I). We measured this index following 
the narrow definition introduced by \citet{balogh99}, with the
continuum bands at 3850--3950 and 4000--4100 \AA; this narrow index will be
referred as $D_n(4000)$. Fig.~\ref{fig_cumulative_colour_CI}c shows the
distribution of this index for our spectral classes.
We clearly note the bimodal distribution of this parameter, with a
divisory line at $D_n(4000) = 1.67$ separating star-forming and
passive galaxies with the highest values of both reliability and  
completeness ($> 98$ per cent).

\begin{figure*}
\resizebox{\textwidth}{!}{\includegraphics{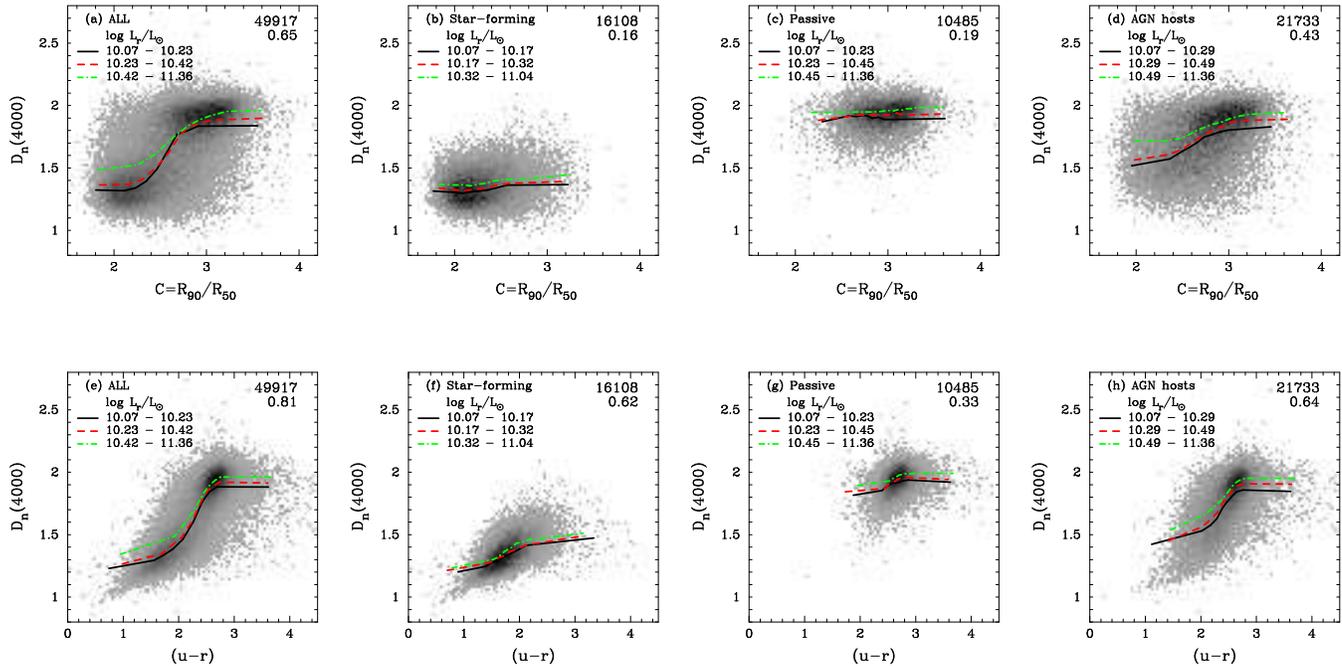}}
\caption{Concentration index and $(u-r)$ colour versus $D_n(4000)$ for all 
galaxies in our sample and for our distinct spectral classes. 
It is also shown the number of galaxies in each panel 
as well as the Spearman rank correlation coefficient ($r_S$).
The median values of $D_n(4000)$ in bins of $C$ and $(u-r)$ for three
ranges of galaxy luminosity are also shown as lines with different shapes.}
\label{CIcolour_D4000}
\end{figure*}

In Fig.~\ref{CIcolour_D4000} we show the $D_n(4000)$ index as a function of
both concentration index and $(u-r)$ colour for all galaxies in our
sample, and for the spectral classes analysed here. The median values
of $D_n(4000)$ in bins of $C$ and $(u-r)$ for three ranges of galaxy
luminosity are also shown as lines with different shapes.
When all galaxies are considered, the $D_n(4000)$ index shows 
a correlation with both concentration index ($r_S = 0.65$) and colour 
($r_S = 0.81$). For star-forming galaxies there is a significant 
correlation only with colour ($r_S = 0.63$), implying that these two
quantities are linked for this type of galaxies. A natural 
explanation for these relations is that young and blue stellar 
populations are responsible for maintaining the correlation among colour
and $D_n(4000)$ for star-forming galaxies. On the other hand, as the 
concentration index is more associated with the shape of a
galaxy than with its stellar population properties, the 
absence of correlation among $C$ and $D_n(4000)$ index reflects that 
star-forming galaxies with  same mean stellar age (or $D_n(4000)$ value) 
span a wide range of morphological types (or concentration values).
We also note in Fig.~\ref{CIcolour_D4000} that passive galaxies do not 
show significant correlations among the quantities analysed here,
and galaxies hosting AGN show correlations in both colour and 
$C$ versus $D_n(4000)$ plots, as expected due to the mixture of
populations comprised by this spectral class.

We thus conclude that by using spectroscopic information
(emission lines and spectral features) one can obtain a better 
splitting between the star-forming and passive galaxy populations
than by using concentration and colours. Our results also indicate 
that galaxies hosting AGNs tend to present intermediate behaviour
with respect to these two populations. Additionally, hereafter we
will consider the star-forming and passive galaxies as the extremes of 
both blue and red galaxy distributions. Moreover, the optimal values that
distinguish these spectral classes, mainly that for the $D_n(4000)$ index,
will be used to define the two main galaxy populations which inhabit
the local universe, historically the early and late galaxy types.

\begin{figure*}
\centerline{\includegraphics[scale=1.0]{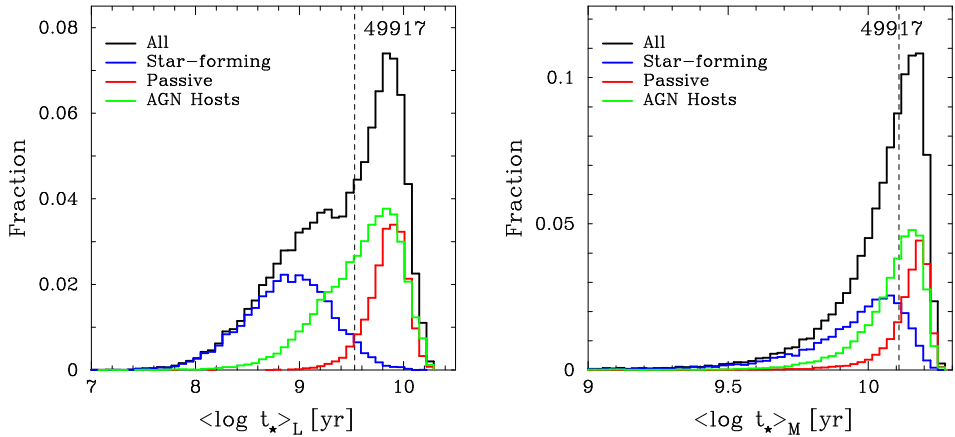}}
\caption{Distribution of mean stellar age weighted by light (left) and by
mass (right) for all galaxy sample, star-forming, passive and AGN hosts
spectral classes. The vertical dashed line in the left panel is the
optimal separator between the star-forming and passive galaxy
distributions.}
\label{fig_age_histogram}
\end{figure*}

\subsection{Physical properties of spectral classes}
\label{sec:physics_classes}

Here we analyse the properties of the main physical parameters
derived from our spectral synthesis method for each spectral class
defined above.

\subsubsection{Mean stellar age}\label{sec:age}

The mean light-weighted stellar age of a galaxy reflects the epoch
of formation of massive and bright O and B stars, frequently associated 
to starbursts. Thus, it is strongly affected by the recent star formation
history of a given galaxy. On the other hand, the mean mass-weighted
stellar age is associated with the epoch of formation of the stellar
population which nowadays contributes most significantly to the galaxy
mass. Consequently, it is associated with the mass assembly history of a
given galaxy.

In Fig.~\ref{fig_age_histogram} we show the distributions of these two age
estimates for all galaxies in our sample, and for each spectral class
separately. The distribution for the mean light-weighted stellar age shows
a conspicuous bimodality with the star-forming and passive galaxies at
their extremes, and the spectral class of AGN hosts occupying an
intermediate locus in it. For this distribution,
$\langle\log t_\star\rangle_L \,\simeq 9.53$ is
the age which best divides the extreme galaxy populations (shown in
Fig.~\ref{fig_age_histogram} as a vertical dashed line). 
On the other hand, the distribution of $\langle\log t_\star\rangle_M$ is
single peaked at older ages, with median value of about 12.4 Gyr for all
sample. This implies that galaxies of all spectral classes in our sample
have a large fraction of their stellar masses in old stellar populations.
Thus, it is seen that the bimodality noted in some galaxy properties, as
discussed above, is primarily related to `light-weighted' quantities. In
other words, recent episodes of star formation present in star-forming
galaxies, in contrast with the absence of significant amount of young
stars in passive galaxies, produce the bimodal distribution seen in the
$\langle\log t_\star\rangle_L$ distribution, as well as in the $(u-r)$
colour and $D_n(4000)$ index distributions.

We have also investigated the behaviour of these trends for the results obtained
with a base excluding the low-$Z$ SSPs, which, as shown in Fig.~\ref{fig_Dn4000_x_SSPs}, tend to result in older ages for young stellar populations. Actually, we found this is true: the values of the mean stellar ages obtained by using the base without the 0.005 and 0.02 $Z_\odot$ SSPs, compared to that obtained with the base adopted in this work, decrease about 0.18 dex when all galaxies are considered and 0.31 dex for star-forming galaxies. The age with best divide the galaxy populations for this `metallic' base is $\langle\log t_\star\rangle_L \,\simeq 9.32$. We also note that the distribution of $\langle\log t_\star\rangle_M$ for this base shows a second peak at younger ages, besides that at older ages. As discussed in Section \ref{sec:Synthesis_method}, this is a consequence of using $Z \ge 0.2 Z_\odot$ in the spectral base, which imply that one has to assume that younger galaxies were formed less than $\sim 1$ Gyr ago, contrarily to the assumption of more continuous star-formation regimes generally invoked to explain the properties of these galaxies.

\begin{figure}
\centerline{\includegraphics[scale=1.0]{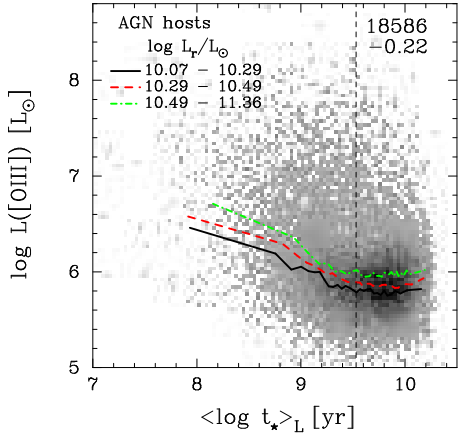}}
\caption{Luminosity of \Oiii\ emission line as a function of the mean
light-weighted stellar age. The median values of $L\oiii$ as a
function of age for three ranges in galaxy luminosity are shown as 
lines with different shapes. The vertical dashed line is the age value
with best distinguish star-forming and passive galaxies.}
\label{fig_AGN_ages}
\end{figure}

As shown in Fig.~\ref{fig_age_histogram}, AGN hosts constitute an intermediate population of galaxies. We analyse this behaviour of AGN hosts by investigating the distribution
of the \Oiii\ luminosity, $L\oiii$ (thought to be a tracer of the AGN
power, e.g. \citealt{kauffmannAGN}), as a function of the mean
light-weighted stellar age. This relation is shown in
Fig.~\ref{fig_AGN_ages}, where we also plot the median values of
$L\oiii$ as a function of $\langle\log t_\star\rangle_L$ for three
ranges in galaxy luminosity (shown as lines with different shapes); the
age value which best separates star-forming and passive galaxies is
also shown as a vertical dashed line. The correlation among these
quantities is in the sense that the $\oiii$ luminosity is low in older
galaxies, with median values larger for brighter galaxies. In galaxies
hosting AGNs with $\langle\log t_\star\rangle_L > 9.53$ the $L\oiii$ is
almost constant, as indicated by its median values in all galaxy
luminosity bins. AGN hosts with younger stellar populations have the
\oiii\ luminosity decreasing with age \citep[see also][for similar
results]{kauffmannAGN}. In other words, Fig.~\ref{fig_AGN_ages} shows
that even AGN hosts have a bimodal distribution with old galaxies
having \oiii\ luminosities lower and constant along age bins (but
increasing with galaxy luminosity), and galaxies with young stellar
populations having high values of $L\oiii$, which increases as a galaxy
becomes younger and brighter.

\begin{figure}
\centerline{\includegraphics[scale=1.0]{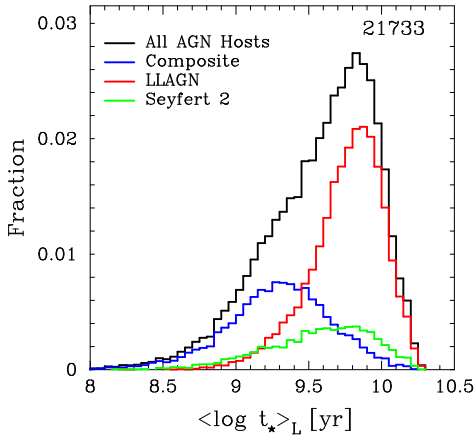}}
\caption{Distribution of mean light-weighted stellar age for distinct
types of AGN hosts (see text for details).}
\label{fig_AGN_histo}
\end{figure}

In order to investigate this apparent bimodal character of AGN hosts, we
have divided this class in three subclasses according to their emission
line properties. The approach used here is analogous to that of
\citet{brinchmann04}. The subclasses are: (i) `composite' galaxies,
which appear in the BPT diagram between the curves defined by
\citet{kauffmannAGN} and \citet{kewley01}; (ii) hosts of `Seyfert 2'
nuclei types, located in the BPT diagram above the curve defined by
Kewley et al.; and (iii) hosts of low-luminosity AGN (LLAGN), which do
not appear in this diagram as they have only the \nii\ and \Ha emission
lines measured with $S/N > 3$, and are classified by using only the
criterion $\log (\nii/\Ha) > -0.2$ \citep[see][]{miller03}.
In Fig.~\ref{fig_AGN_histo} we show
the distribution of light-weighted mean stellar ages for these three
subclasses of AGN hosts, as well as for all galaxies in this spectral
class. It is interesting to note that composite galaxies at one side,
and LLAGN hosts at another, are in the tails of the distribution for
all AGN hosts. The majority of AGN hosts in our sample is composed
by galaxies with old stellar populations, mainly represented by the
LLAGN hosts which show low values for the \Oiii\ luminosity.
We conclude from the $L\oiii$ versus age relation shown in Fig.\ref{fig_AGN_ages}
and the distributions seen in Fig.~\ref{fig_AGN_histo} that the AGN activity
is closely linked to the star-formation activity of a galaxy, in the sense
that galaxies with younger stellar populations tend to have more powerful AGNs.

\subsubsection{Stellar mass}

\begin{figure}
\centerline{\includegraphics[scale=1.0]{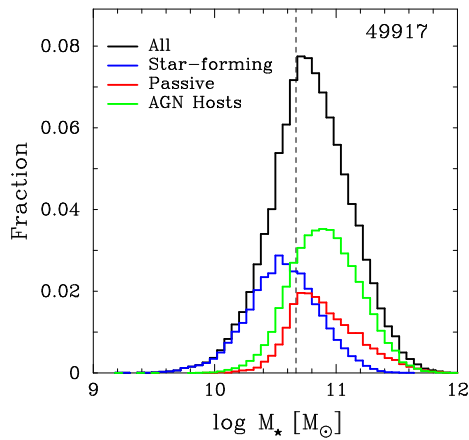}}
\caption{Distribution of stellar mass for all galaxies in our sample, 
and for classes of star-forming, passive and AGN hosts. 
The vertical line is the optimal separator between star-forming and 
passive galaxies.}
\label{fig_mass_histogram}
\end{figure}

The distribution of stellar mass for our sample is presented in
Fig.~\ref{fig_mass_histogram}. As in Fig.~\ref{fig_age_histogram}, we
show the distributions for the full sample and for each spectral class
discussed here. The value of stellar mass which best divides
star-forming and passive galaxies is shown as a vertical dashed line,
and corresponds to $M_\star \sim 4.7 \times 10^{10}$~\msun.  The
reliability and completeness parameters obtained with these values are
$\sim 71$ per cent. 

This value is susbtantially larger than the $3
\times 10^{10}$~\msun\ at which \citet{kauffmannII} find a sharp
transition in the physical properties of galaxies. As mentioned in SEAGal I, this
difference cannot be attributed to the differences in IMF.  Kauffmann
et al.\ used the Kroupa IMF, whereas we use the one by Chabrier, but
as show in the BC03 paper, these two IMFs yield identical $M/L$
ratios.

A more relevant source of discrepancy is related to the approach used
to determine the transition mass. In \citet{kauffmannII}, there is no
objective criteria to define the stellar mass value at which galaxy
properties change, whereas we have used a clear procedure to find the
stellar mass transition at $M_\star \sim 4.7 \times
10^{10}$~\msun.

We have also verified if the use of a volume-limited sample in such analysis,
with an absolute magnitude limit cutoff $M(r) = -20.5$, affects the determination of
the transition mass value, being responsible for the difference
discussed above.  In order to investigate this issue we have
determined the transition mass for galaxies in a flux-limited sample
containing 20000 galaxies selected at random from the main galaxy
sample of SDSS. This results in an even larger transition mass,
$M_\star \sim 6.3 \times 10^{10}$~\msun.

As anticipated in section \ref{sec:Synthesis_method}, the inclusion of
very low $Z$ SSPs in the base inevitably leads to larger stellar
masses.  The \citet{kauffmannII} mass estimates are based on a library
of model galaxies constructed with $Z \ge 0.25 Z_\odot$, whereas the
SSPs in our base go down to $0.005 Z_\odot$. This is a systematic
source of discrepancy between the mass scales in these two studies.
In SEAGal I, which used only $Z \ge 0.2 Z_\odot$ SSPs, we have shown
that the stellar masses estimated via the spectral synthesis approach
are about 0.1 dex larger than that obtained by Kauffmann et
al. (MPA/JHU group).  Repeating this comparison for our new stellar
masses yields a median offset of 0.2 dex. As discussed in SEAGal I,
part of this difference ($\sim 0.1$ dex) could be attributed to
technical details employed to estimate the stellar masses. The
remaining difference is due to the inclusion of low-$Z$ SSPs in our
spectral base. We have confirmed this hypothesis through the analysis
of the results obtained with a base excluding the 0.002 and 0.005
$Z_\odot$ SSPs. As expected, the $M_\star$ values obtained by this way
are only about 0.1 dex larger than the stellar masses obtained by the
MPA/JHU group. The transition mass for this base is $M_\star \sim 3.9 \times 10^{10}$~\msun.

Therefore, discrepancy between the values of the stellar mass
transition obtained here and in \citet{kauffmannII} is due to a
combination of differences in the methodology to define this mass and
differences in the metallicities in reference set of model stellar
populations.

\begin{figure*}
\resizebox{\textwidth}{!}{\includegraphics{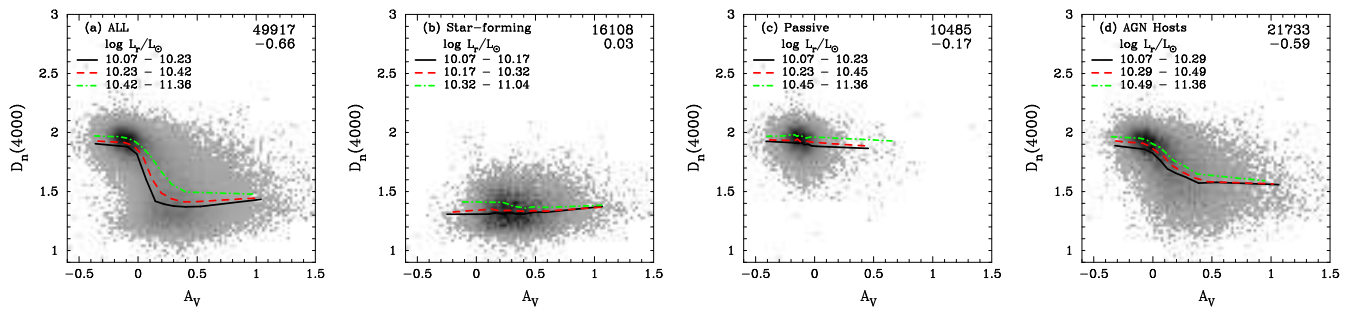}}
\caption{$V$-band stellar extinction versus $D_n(4000)$ for (a) all
galaxy sample, (b) star-forming, (c) passive, and (d) AGN host
galaxies. The median values of $A_V^\star$ in bins of $D_n(4000)$
index for three luminosity intervals are shown as different lines.}
\label{fig_AV_D4000}
\end{figure*}

\subsubsection{Stellar extinction}

Another important quantity related to physical processes occurring in
galaxies is the amount of dust present in the interstellar medium.
Our spectral synthesis approach estimates this value by obtaining the
attenuation of starlight by dust parametrised in $V$-band, i.e. the
stellar extinction, $A_V^\star$.  We plot in Fig.~\ref{fig_AV_D4000}
this parameter against the $D_n(4000)$ index. The results for all
galaxies in our sample are depicted in Fig.~\ref{fig_AV_D4000}a. There
is a clear anti-correlation between $A_V^\star$ and $D_n(4000)$ with
$r_S=-0.66$. However, the relation between these two quantities is
better appreciated when we consider separately star-forming, passive
and AGN host galaxies.  Indeed, star-forming galaxies
(Fig.~\ref{fig_AV_D4000}b) do not show significant correlation between
$A_V^\star$ and $D_n(4000)$, with $r_S=0.03$, and passive galaxies
(Fig.~\ref{fig_AV_D4000}c) exhibit only a small anti-correlation
($r_S=-0.17$).  The median of $A_V^\star$ for star-forming and passive
galaxies are 0.33 and -0.12, respectively.

The relation between $A_V^\star$ and $D_n(4000)$ is weak for 
star-forming and passive galaxies, but strong for galaxies
harbouring an AGN, as can be appreciated in  Fig.~\ref{fig_AV_D4000}d.
The Spearman  correlation coefficient between these
two quantities is $-0.60$. It is not easy to interpret this finding,
considering that the measured spectrum of this kind of object is a 
composite of the AGN spectrum plus the underlying stellar population. 
A better understanding of what is going on requires spatially-resolved
spectroscopy of the central regions of this kind of galaxy.  

\begin{figure*}
\resizebox{\textwidth}{!}{\includegraphics{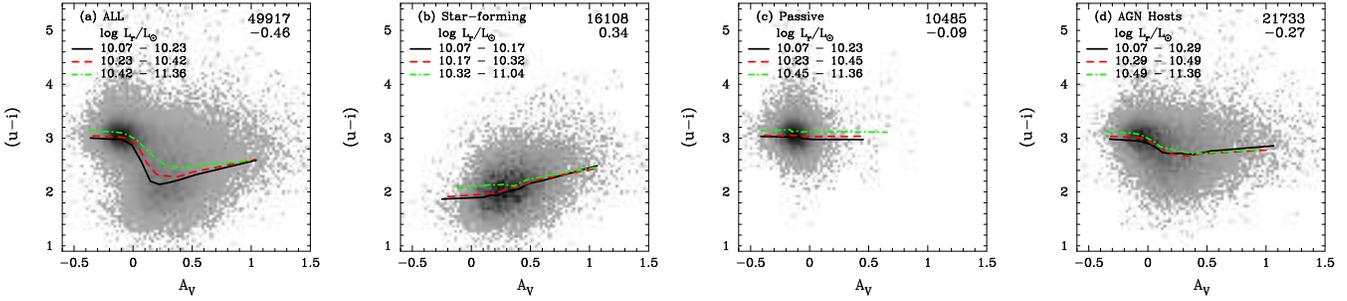}}
\caption{Same as Fig.~\ref{fig_AV_D4000}, but now with $(u-i)$ colour 
in the ordinate.}
\label{fig_AV_colour}
\end{figure*}
 
It is interesting to verify how the trends above behave if we use 
galaxy colours. This is shown in Fig.~\ref{fig_AV_colour} for 
the colour $(u-i)$. The trends of stellar extinction with colour are, 
qualitatively, quite similar to those between $A_V^\star$ and 
$D_n(4000)$, when all galaxies are considered. However, in the case of 
star-forming galaxies, the correlation become stronger with 
colour (the Spearman coefficient being $0.33$). This is in agreement
with the results reported by \citet{stasinska04} for the nebular 
extinction of spiral galaxies: it tends to increase with galaxy colour.
On the other hand, the small correlation shown in 
Fig.~\ref{fig_AV_D4000}c for passive galaxies disappear in the plot
of $A_V^\star$ versus $(u-i)$ colour. We discuss other properties of 
extinction in another paper of this series \citep{sodre06}.

\subsubsection{A note about negative extinction}

As discussed in SEAGal I, for both statistical and physical reasons, we did
not constrain the extinction $A_V^\star$ to be positive. Interestingly,
$A_V^\star < 0$ galaxies are nearly all passive, in agreement with
\citet{kauffmannI}, who found negative extinction primarily in galaxies with a
large $D_n(4000)$. As a whole, the distribution of $A_V^\star$ for passive
galaxies indicates that they have essentially no dust, as expected given that
they are dominated by old stellar populations. Still, the fact that this
distribution is centered on slightly negative values of $A_V^\star$ deserves
some explaining, as it cannot be attributed purely to random errors, which
would produce $A_V^\star = 0$ on average.  Although this has no impact on the
main conclusions of this paper, we find it useful to open a parenthesis to
look at this issue more closely.

We find that one of the factors which is responsible for this tendency is
$\alpha$-enhancement. As reported by \citet{sodre05_cancun}, our STARLIGHT
fits for passive galaxies underpredict the strength
of $\alpha$-element features, which is not surprising given that the
STELIB-based BC03 SSPs do not account for the $\alpha$-enhancement known to
occur in massive ellipticals \citep{worthey92,davies93,thomas02}.
Furthermore, the residuals in these
features are larger as $\sigma_\star$ increases, in analogy with the
[Mg/Fe]--$\sigma_\star$ anti-correlation.

Gomes (2005) carried out simulations to test the effects of enhancing
$\alpha$-features in test galaxy spectra constructed with the BC03 SSPs. To
emulate this effect realistically, he has added to the models the mean
residual spectrum of passive galaxies in the regions of CN, Mg and NaD. These
controlled experiments showed that in order to try to match these artificially
enhanced bands, STARLIGHT overpredicts the strength of the older components,
where these features are deeper. Furthermore, even though the test galaxies
were constructed with $A_V^\star = 0$, the fits yield systematically negative
extinction, of order $A_V^\star \sim -0.1$, as found in passive galaxies
(Fig.~\ref{fig_AV_D4000}b). This happens because the older components invoked
to match the deeper $\alpha$-spectral features are also redder, implying a
continuum mismatch which is compensated by a slightly negative $A_V^\star$.

The bottom line is that negative extinctions in passive galaxies are at least
partly due to the use of a base of SSPs which do not have the proper chemical
mixture. Recently published $\alpha$-enhanced libraries should help fixing
this mismatch \citep[e.g.][]{coelho05,munari05}.


\section{Discussion}\label{sec:Discussion}

\subsection{Origin of the bimodal distribution}
\label{sec:origins_bimodality}

\begin{figure*}
\resizebox{\textwidth}{!}{\includegraphics[scale=1.0]{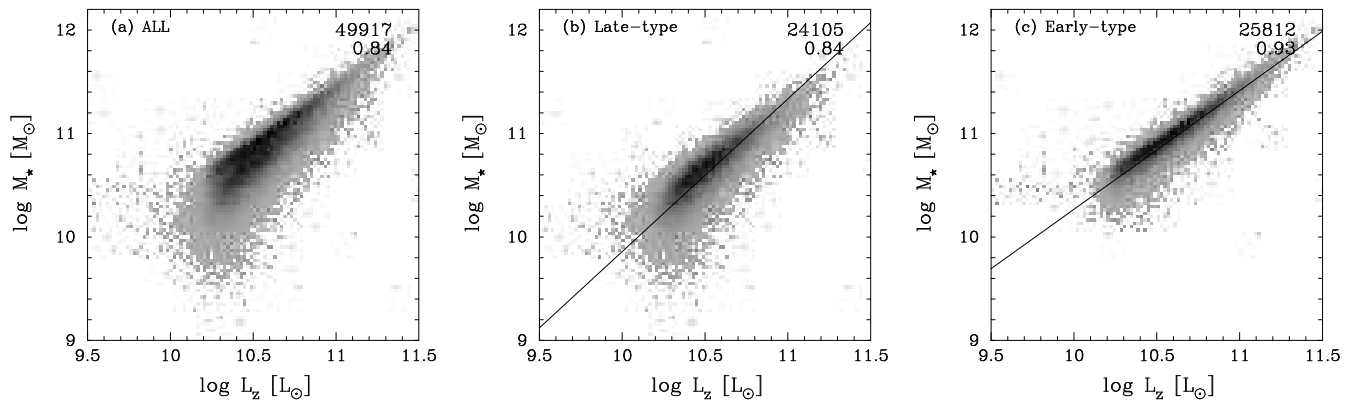}}
\caption{Mass-luminosity relation for (a) all galaxies in our sample,
and for (b) late-type galaxies with $D_n(4000) < 1.67$ and (c)
early-type galaxies with $D_n(4000) > 1.67$. Solid lines are robust
fits for the relations.}
\label{fig_mass_luminosity}
\end{figure*}

An easy way to visualise the bimodality of the galaxy population is in 
the colour-magnitude diagram, which shows two separate colour sequences,
a broad blue sequence, and a narrower red one
\citep[see e.g.][]{baldry04}. Here we investigate the nature of bimodality through
the mass-luminosity relation. Fig.~\ref{fig_mass_luminosity} shows the 
stellar mass as a function of the dust-corrected $z$-band luminosity 
(in solar units) for the full galaxy sample and for the galaxy populations
divided by using the value $D_n(4000) = 1.67$. Robust linear fits to the
mass-luminosity relation for these two galaxy sequences yield:
\[
\log (M_\star/M_\odot) = -4.89 + 1.48\, \log (L_z/L_\odot)\textrm{, for late-types}
\]
and
\[
\log (M_\star/M_\odot) = -1.50 + 1.18\, \log (L_z/L_\odot)\textrm{, for early-types.}
\]
Here, the two
galaxy sequences commonly seen in the colour-magnitude diagram are well
characterised with that corresponding to early-type galaxies being less
steep than that of late-type galaxies. The relation for early-types also shown
the best correlation coefficient ($r_S = 0.93$) and less scatter. For late-type galaxies,
the scattering observed in the $M_\star - L_z$ relation is associated to the large spread in the mass-to-light ratio and star formation histories shown by these objects.

Recently, \citet{baldry04} have found that a change in the 
distribution of galaxy colours occurs around
$(1-3) \times 10^{10}$~\msun, close to the value at which the galaxy
properties also change \citep[see][]{kauffmannII}.
However, from Fig.~\ref{fig_mass_luminosity} we note that another parameter is needed in order to achieve a better separation between the two galaxy populations
characterised by the extreme classes of star-forming and passive galaxies.
For instance, the value
$\log M_\star/M_\odot > 10.5$ can select almost all early-type galaxies,
but a large number of massive late-type galaxies is also selected by
using this criterion. In fact, as shown in Table \ref{completeness},
the reliability and completeness parameters for
star-forming and passive galaxies obtained with the optimal value of
$\log M_\star/M_\odot = 10.67$ are lower than those obtained when
considering the mean light-weighted stellar age to split the galaxy
populations. Thus, it seems that the galaxy mass is not playing a main
role in defining the bimodal galaxy distribution seen in local galaxies,
reinforcing the idea that the observed bimodality is related to the
presence of a young component in low-mass, blue galaxies.

\begin{figure*}
\centerline{\includegraphics[scale=0.75]{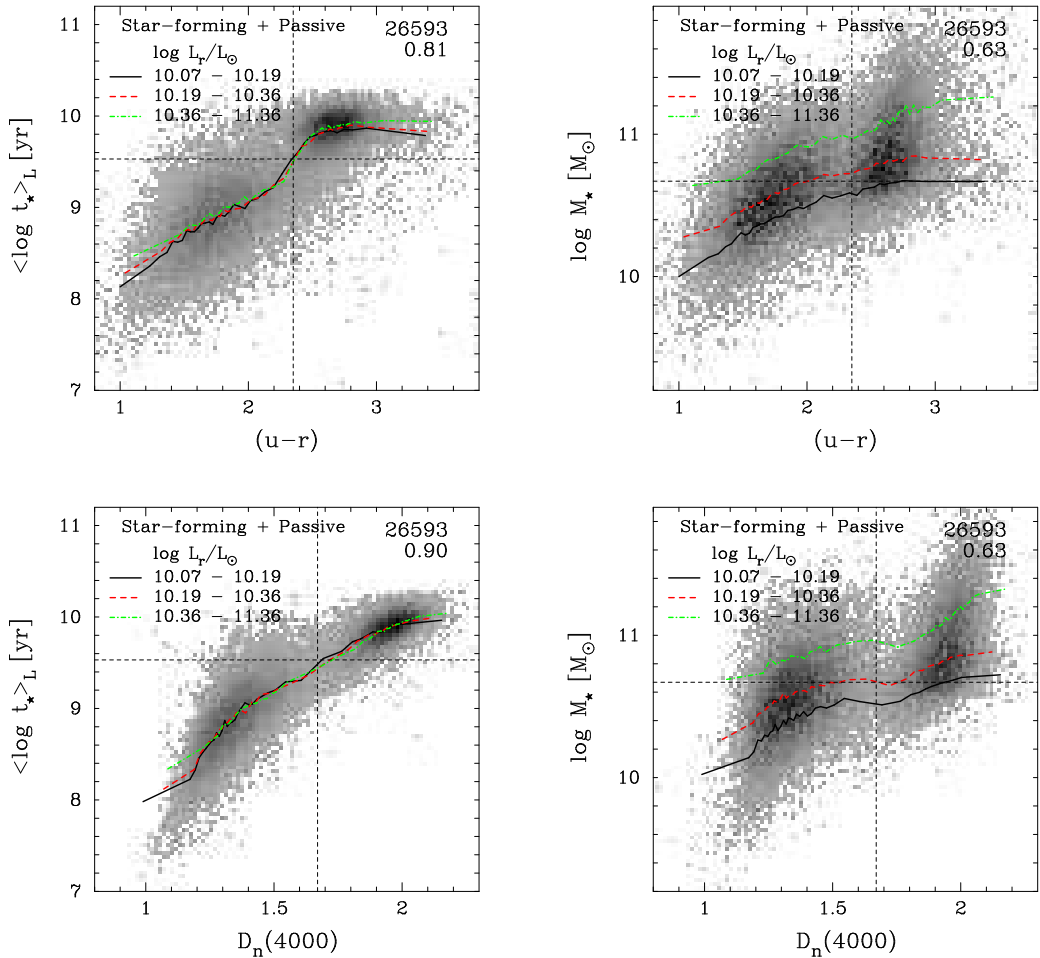}}
\caption{Mean stellar age and stellar mass as a function of $(u-r)$
colour (top panels) and $D_n(4000)$ index (bottom panels), for star
forming and passive galaxies in our sample. It is also shown the number
of objects and the Spearman rank correlation coefficient for each
panel.}
\label{fig_colour_D4000}
\end{figure*}

Another way to investigate this issue is presented in 
Fig.~\ref{fig_colour_D4000}, where we show the $(u-r)$ colour
and $D_n(4000)$ index versus the mean light-weighted stellar age and
stellar mass for star-forming and passive galaxies in our sample. 
The values which better divide these extreme galaxy populations are 
shown as vertical and horizontal dashed lines. The median values of
colour and $D_n(4000)$ for three bins of galaxy luminosity are also
shown as different lines.

With the help of this figure, we have investigated the trend of
the mean light-weighted stellar age to distinguish 
the extreme galaxy populations in a better way than if using stellar mass.
Considering the optimal values of $(u-r)$ colour and 
$D_n(4000)$ index, and that of \tl\ and $\log M_\star$, to characterise
early and late-type galaxies, there are two sets of \textit{abnormal}
galaxies (located in the top-left and bottom-right quadrants of 
Fig.~\ref{fig_colour_D4000}): 
late-type galaxies (e.g. with $D_n(4000) < 1.67$) with old stellar
ages or high stellar masses, and early-type galaxies (with
$D_n(4000) > 1.67$) with young stellar ages or low stellar masses. We
find that the fraction of late-type `old' plus that of early-type
`young' galaxies is about 6 per cent, whereas the fraction of
late-type `massive' plus early-type 'low-mass' galaxies is 29 per
cent. Thus, taking a single value to characterise the two main
galaxy populations (star-forming and passive), for instance based on 
the $D_n(4000)$ index, the fraction of those uncommon galaxies is 
higher when we consider the optimal value of the stellar mass, in 
contrast with the fraction obtained when considering the mean 
stellar age to split the galaxy population. A similar trend 
is also seen when all galaxies are taken into account.

Thus, supported by these findings, we argue that the bimodality of the
galaxy population, commonly seen in colour-magnitude diagrams, single 
colour distributions, and in the mass-luminosity relation discussed 
above, is related to the presence of a young and luminous stellar
component in galaxies currently undergoing star formation, in contrast
with the stellar content of passive galaxies, where old stars are the
responsible for most of their luminosities.

\begin{center}
\begin{table*}
\begin{tabular}{|c|c|c|c|c|c|c|}
  \hline
 ~ & Optimal value & $\mathcal R_{\rm SF}$  & $\mathcal R_{\rm P}$ & $\mathcal C_{\rm
SF}$  & $\mathcal C_{\rm P}$ & $\mathcal{C_{\rm SF} R_{\rm SF} C_{\rm P} R_{\rm
P}}$\\
  \hline
$C$                   & 2.62 & 86.0 & 89.8 & 93.0 & 80.3 & 57.6 \\
$(u-r)$               & 2.35 & 93.6 & 93.8 & 96.0 & 90.4 & 76.2 \\
$(u-i)$               & 2.68 & 91.7 & 95.2 & 96.7 & 88.2 & 74.4 \\
$D_n(4000)$           & 1.67 & 98.8 & 98.2 & 98.6 & 98.1 & 93.9 \\
\tl                   & 9.53 & 94.8 & 92.9 & 95.4 & 92.0 & 77.3 \\
log $M_\star/M_\odot$ &10.67 & 64.2 & 79.9 & 83.0 & 59.2 & 25.2 \\
  \hline
\end{tabular}
\caption{Reliability and completeness parameters for concentration index,
$(u-r)$ and $(u-i)$ colours, $D_n(4000)$ index, mean light-weighted stellar age and
stellar mass optimal values obtained to split the distributions of
star-forming and passive galaxies.}
\label{completeness}
\end{table*}
\end{center}

\subsection{The role of stellar mass}

Colours of galaxies reflect their star formation histories, which are 
also linked to the stellar populations that we are seeing today. 
For instance, a galaxy that has suffered recent bursts of star formation
will contain a considerable fraction of young and hot stars that
will be responsible for a large fraction of the light we receive 
from it. On the other hand, earlier events of huge star formation 
activity will keep the fraction of old stellar populations even higher,
leading to galaxies with redder colours and spectra dominated by 
absorption lines. 

We also observe that red, passive galaxies tend to be more massive
than blue, star-forming galaxies. Thus, the stellar mass of a galaxy 
is related to the bimodality seen in the colour distribution of local
galaxies, in the sense that only less massive galaxies appear to be
forming stars in the last few gigayears.

\begin{figure*}
\resizebox{\textwidth}{!}{\includegraphics[scale=1.0]{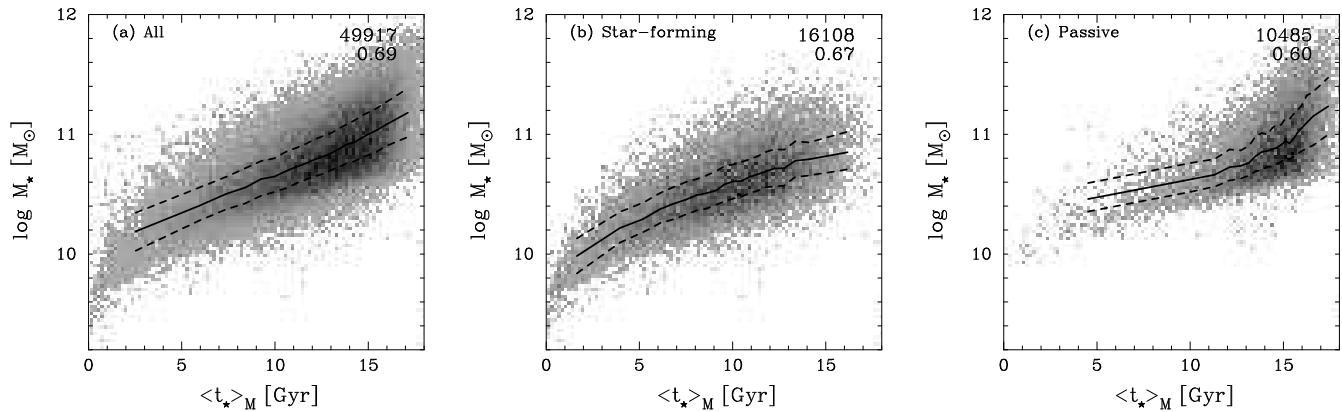}}
\caption{Stellar mass versus the mass-weighted mean stellar age of
formation of the stellar population which more contributes in mass,
for (a) all galaxy sample, (b) star-forming, and (c) passive galaxies.
The solid lines are the median and quartile values of stellar mass in 
age bins containing the same number of galaxies.}
\label{fig_lookback_age_mass}
\end{figure*}

We have investigated this tendency by examining the
mass-weighted mean stellar age, $\langle t_\star\rangle_M$
\footnote{actually $\langle t_\star\rangle_M = 10^{\langle\log
t_\star\rangle_M}$}, of galaxies in our sample, as a function of their
stellar masses. This age is related to the epoch of formation of the
stellar population which nowadays contributes significantly to the
galaxy mass. In Fig.~\ref{fig_lookback_age_mass} we show the relation between
stellar mass and $\langle t_\star\rangle_M$ (in Gyr), for (a) all galaxy
sample, (b) star-forming, and (c) passive galaxies.
There is a clear correlation between stellar mass and
$\langle t_\star\rangle_M$, with a Spearman rank coefficient of
$r_S=0.69$ for all galaxy sample. The median values of stellar mass in
bins of stellar age containing the same number of objects are also
shown in this figure. Note that the median stellar mass increase from
$1.4 \times 10^{10}$ to $1.5 \times 10^{11}$ M$_\odot$, from 
the young to the oldest age bin. This trend is particularly related 
to the existence of a dominant fraction of passive galaxies, with high
stellar masses and without signs of recent star formation activity,
at the oldest age bins shown in Fig.~\ref{fig_lookback_age_mass}c.
In addition, less massive galaxies are still forming stars which
will contribute to increase their masses further.

\begin{figure*}
\centering
\includegraphics[scale=1.0]{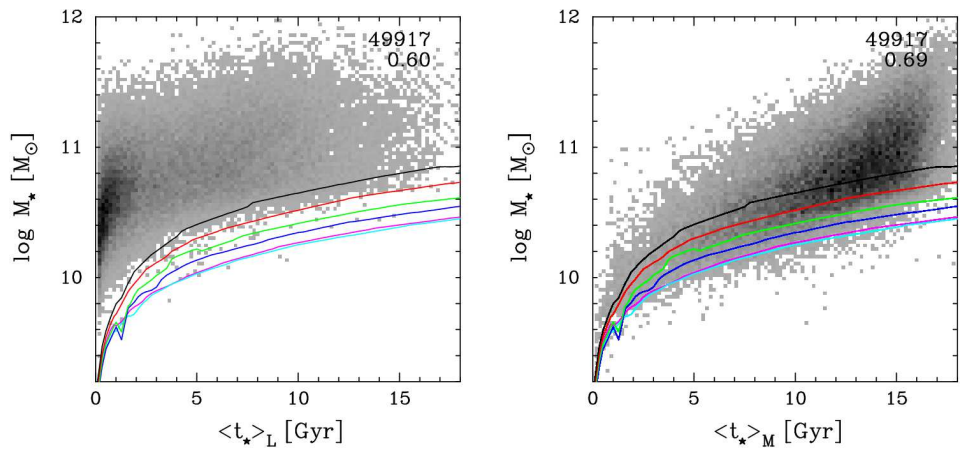}
\caption{Stellar mass as a function of the mean light-weighted stellar
age (left panel) and mass-weighted stellar age (right panel). The solid
lines represent the stellar mass for which a SSP of age
$\langle t_\star\rangle_L$ has an absolute $r$-band magnitude of
$-20.5$. Each line corresponds to a stellar metallicity. From bottom
to top: $Z=0.005, 0.02, 0.2, 0.4, 1$ and 2.5 $Z_\odot$.}
\label{fig_testing_mag_limit}
\end{figure*}

When examining these plots, it is fit to recall
that we are using a volume-limited sample. The limiting magnitude of
$M(r) = -20.5$ (corresponding to $M^*(r) + 1$) is large enough to
detect massive galaxies of any age, but old low-mass systems will be
excluded by construction.  To examine this issue quantitatively, we use
the SSPs in BC03 and determine, for each age and metallicity, the
stellar mass which results in $M(r) = -20.5$. The results are shown as
solid lines, one for each $Z$, in Fig.~\ref{fig_testing_mag_limit},
overplotted to the $M_\star$ versus $\langle t_\star\rangle_L$ and
$M_\star$ versus $\langle t_\star\rangle_M$
relations for our sample. The $M_\star(t,Z)_{M(r) = -20.5}$ lines trace
very well the lower envelope of the data points in the left panel of Fig.~\ref{fig_testing_mag_limit}. The
lower envelop becomes fuzzier using $\langle t_\star\rangle_M$, but
the overall conclusion is still the same. Except for a handful of cases,
all galaxies in this sample lie above these lines. Old, low mass objects
(``dwarf elliptical''-like) which would populate the lower-right region of this
diagram are too faint to satisfy our $M(r) < -20.5$ cut. On the other
hand, and perhaps more interestingly, the nonexistence of massive
galaxies with young stellar populations (mainly noted in the relation with
$\langle t_\star\rangle_M$), which should appear at the
top-left corner of Fig.~\ref{fig_testing_mag_limit}, is clearly
unrelated to our sample selection. Thus, our claim that massive
galaxies have essentially older stellar populations whereas, in
contrast, less massive ones have younger populations, does not depend
on the limiting magnitude adopted in this work.

\subsection{Downsizing in galaxy formation}

In a biased galaxy formation scenario \citep[e.g.][]{cen93}, massive
galaxies form from the highest peaks in the initial density
fluctuation field. In this scenario, massive galaxies formed at
high redshifts tend to inhabit the high-density regions associated
to galaxy clusters and rich groups seen today. Indeed, \citet{kodama04} 
in a photometric analysis of galaxies in
colour-selected high-density regions at $z \sim 1$,
have shown that galaxy formation processes, including mass
assembly and star formation, take place rapidly and are completed
early in massive systems, while in the less massive ones these
processes (at least star formation) are slower.

This galaxy formation scenario, in which more massive galaxies are
formed at higher redshift, was first proposed
by \citet{cowie96}, who found that the maximum luminosity (or
mass) of galaxies undergoing rapid star formation has been
declining smoothly with decreasing redshift since $z \sim 1$. 
Recently, many observational results have supported this idea by suggesting
that the most luminous galaxies formed the bulk of their stars in the first
$\sim 3$ Gyr of cosmic history
\citep[][and others]{mccarthy04,kodama04,juneau05}.

Hence, even at $z \sim 1$ there would be a red and luminous population
composed by massive and old galaxies, and a blue and faint one,
formed by less massive galaxies with still ongoing star formation.
Recently, evidence for this bimodal distribution of galaxy
properties (at least in colour) around this redshift has been
found by many works through different approaches and using different
observational data sets \citep{bell04,wiegert04,weiner05}.

In the local universe, the bimodality of galaxy properties was
firstly noted in the colour distribution of SDSS galaxies 
\citep[e.g.][]{strateva01} and in the star formation properties of
galaxies from the 2dFGRS \citep[e.g.][]{madgwick02}. In the last
years many works have focused their analysis on the dependence of
galaxy properties on the stellar mass. For instance, 
\citet{kauffmannII} have shown that at a stellar mass above $3 \times
10^{10}$~M$_\odot$ there is a rapidly increasing fraction of
galaxies with old stellar populations, high surface mass densities
and high concentrations typical of bulges. These results, and
similar ones based on the star formation properties of galaxies
\citep[e.g.][]{brinchmann04}, together with our own findings derived 
from the analysis of the mass-weighted mean stellar age of galaxies, 
are in perfect agreement with the `downsizing' scenario proposed by
\citet{cowie96}.

Notwithstanding the numerous observational reasons to believe in this 
scenario, it is still very puzzling when we assume the hierarchical 
galaxy formation (bottom-up) scenario of $\Lambda$CDM models, 
which predicts that small systems collapse first and massive galaxies 
form later via the assembly of these already collapsed structures.
The theoretical `downsizing' picture still requires
significant improvements, mainly concerning galaxy formation and 
related processes occurred since $z = 1$, as pointed out by
\citet{hammer05}. At this point it is also important to stress the recent
advance of semi-analytic models based on hierarchical clustering to
explain some observational properties of galaxies, such as the
bimodality in galaxy colours \citep[e.g.][]{menci05}. In this framework,
distinct merging and star formation histories for progenitors with 
different masses lead to a high-$z$ origin of the bimodality seen in 
the properties of galaxies up to $z \sim 1$, particularly during the 
formation and merging of their progenitors.

Despite the uncertainties in the  scenarios for galaxy evolution and the 
current status of  `downsizing' in galaxy formation and hierarchical 
models, here we have shown that our spectral synthesis method discussed in
SEAGal I enables us to investigate the star formation history of
local galaxies, making a link among these objects and their 
progenitors, which have been recently explored by current 
high-redshift galaxy surveys. In future works we will 
discuss in detail this and other related issues.

\section{Summary}
\label{sec:Conclusions}

In this second paper of the SEAGal collaboration, we have investigated 
the bimodality observed in galaxy populations by inspecting the
spectral properties of SDSS galaxies. We have used the physical
parameters derived from the spectral synthesis method applied to a 
sample of about 50 thousand galaxies extracted from the SDSS Data 
Release 2. Galaxies are classified according to their emission line
properties in three distinct groups: star-forming, passive, and AGN
hosts. The bimodality of galaxy properties is investigated with
emphasis on these spectral classes. 
Our main findings are summarised below:

\begin{enumerate}
\item The bimodality of the galaxy population can be represented by
two extreme spectral classes, corresponding to star-forming galaxies
at one side, containing young stellar populations and preferentially of
low stellar masses, and passive galaxies at the other side, without ongoing star
formation and populated by older stars.
\item In an intermediate locus, there are the AGN hosts, which comprise
a mix of young and old stellar populations. However, this spectral
class also shows a bimodal behaviour when considering the \oiii\
emission line luminosity, with younger galaxies showing the larger 
values of $L(\oiii)$, and older ones presenting the same low 
luminosities ($L(\oiii) < 10^6$~L$_\odot$) essentially at all stellar
ages.
\item As a main result, we found that the mean light-weighted stellar age
of galaxies is the direct responsible for the bimodality observed in the
galaxy population.
\item The stellar mass, in this view, has an additional role since most of the star-forming galaxies present in the local universe are low-mass galaxies. Our results also
reinforce the idea of a `downsizing' in galaxy formation, where massive galaxies seen nowadays have stopped to form stars more than 10 Gyr ago.
\end{enumerate}

In this work we have explored a piece of the arsenal of physical parameters
provided by the spectral synthesis method, in order to investigate the
spectral properties of galaxies and revisited the bimodal character of the
galaxy population. Other papers in this series will bring more insight to
some fundamental problems concerning to the physical properties of local
galaxies.


\section*{Acknowledgements}

We thank the anonymous referee for comments and suggestions
that helped improve the paper. We thank financial support from CNPq,
FAPESP and the CAPES/Cofecub program. All the authors also wish to thank
the team of the Sloan Digital Sky Survey for their dedication
to a project which has made the present work possible.

Funding for the Sloan Digital Sky Survey has been provided by the 
Alfred P. Sloan Foundation, the Participating Institutions, the 
National Aeronautics and Space Administration, the National Science 
Foundation, the U.S. Department of Energy, the Japanese 
Monbukagakusho, and the Max Planck Society. 
The SDSS is managed by the Astrophysical Research Consortium (ARC) 
for the Participating Institutions. The Participating Institutions are 
The University of Chicago, Fermilab, the Institute for Advanced Study, 
the Japan Participation Group, The Johns Hopkins University, the 
Korean Scientist Group, Los Alamos National Laboratory, the 
Max-Planck-Institute for Astronomy (MPIA), the Max-Planck-Institute 
for Astrophysics (MPA), New Mexico State University, University of 
Pittsburgh, University of Portsmouth, Princeton University, the 
United States Naval Observatory, and the University of Washington.

\appendix

\section{Aperture effects}\label{appendix:aperture_bias}

A particular effect acting in modern galaxy redshift surveys is
related to the fixed-size apertures used to obtain spectroscopic data.
In the 2dFGRS the fibre diameter is only 2 arcsec, whereas in the SDSS 
the fibres cover 3 arcsec of the sky, leading to a small 
sampling of the integrated light of nearby galaxies.
Recently, \citet*{kewley05} have examined this problem in detail by
investigating the effect of aperture size on the star formation rate,
metallicity, and extinction determinations for galaxies selected from 
the Nearby Field Galaxy Survey. The main result obtained by these 
authors is the need to select galaxies with $z > 0.04$, in 
the case of the SDSS, to minimise aperture effects in the
spectral measurements. This redshift limit is required to assure that
a fibre captures more than 20 per cent of the galaxy light.

In this work, the galaxy sample was built to contain only galaxies 
with redshifts larger than $z = 0.05$. Hence, following the 
recommendation given by Kewley et al., the aperture effects in our 
sample tend to be strongly minimised. On the other hand, in
Fig.~\ref{fig_aperture_effects} we have shown that the fraction of 
star-forming galaxies increases significantly with increasing redshift, 
probably due to the fact that in nearby galaxies the fibre is capturing 
mainly their bulge regions, whereas the nebular emission lines 
(used here to classify galaxies) come essentially from their discs. 
Here, we inspect some aspects of this effect.

\begin{figure}
\centerline{\includegraphics[scale=0.75]{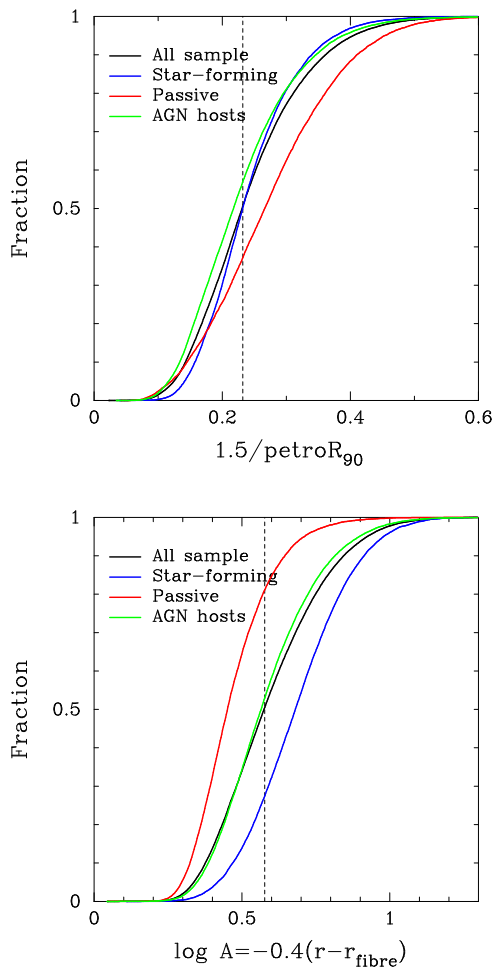}}
\caption{(Top) Cumulative distribution of the ratio of the fibre radius to
the galaxy radius for all sample and for each spectral class discussed in
this work. (Bottom) Cumulative distribution of the correction factor 
to be applied to correct for aperture effect, also for all sample and 
spectral classes. The vertical dashed lines in these plots are the 
median values of the distributions for all sample.}
\label{fig_appendix1}
\end{figure}

In Fig.~\ref{fig_appendix1} we show the cumulative distributions of two
quantities related to the covering fraction of the fixed-size aperture for
all galaxy sample, and for each spectral class discussed in this work.
In the top panel of this figure, it is shown the ratio between the fibre
radius (1.5 arcsec) and the Petrosian radius containing 90 per cent of the
galaxy flux (in $r$-band). Thus, this ratio represents the fraction of the
galaxy size that is actually covered by the fibre. The median value for
all galaxy sample, and also for each spectral class taken individually, is
about 23 per cent. Thus, the fibre covers, on average, the same fraction
of galaxy size almost independently of the galaxy type.

The bottom panel of Fig.~\ref{fig_appendix1} shows the aperture correction,
$A$, to be applied in flux related measurements, based on the
differences between the total galaxy magnitude in the $r$-band and 
the magnitude inside the fibre ($r_{fibre}$), explicitly 
$A=10^{-0.4(r-r_{fibre})}$ \citep{hopkins03}. This correction factor
is commonly applied to emission line derived parameters (like \Ha SFR),
and assumes that the regions where the lines are formed (mainly \hii 
regions) have a distribution identical to that where the continuum is
produced by the old stellar populations. In Fig.~\ref{fig_appendix1} we
note that star-forming galaxies have aperture corrections larger
than that of other classes. This may be somewhat problematic, 
because this kind of correction is performed just in this
type of galaxies \citep[e.g.][]{hopkins03}.

From the plots shown in Fig.~\ref{fig_appendix1}
we conclude that while the 3 arcsec diameter fibres cover about the 
same fraction of galaxy size for all galaxies, the fraction of 
light gathered by their small apertures are biased towards lower 
values (or consequently higher aperture corrections) for star-forming 
galaxies, and higher values for passive galaxies. Since the expected 
aperture effects would tend to increase the number of passive galaxies 
relative to that of star-forming at lower redshifts, the low aperture 
corrections obtained for these galaxies indicate that, on average, 
there are not substantial effects acting on the fraction of this 
spectral class along the redshift range considered in this work. 

\subsection{Mass-to-light ratio}

In order to estimate the total stellar mass of a galaxy from the spectral synthesis, we have assumed that it has a constant radial $M/L$. In other words, total stellar masses are computed from the fibre and Petrosian (total) magnitudes by considering the $M/L$ inside the fibre as representative of the whole galaxy. The effects of this assumption on our stellar mass determination is investigated here.

We have carried out the following tests to address this interesting
question.  From the SDSS photometry data we have both fibre and total
fluxes and colours, from which one can also compute outside-the-fibre
fluxes and colours. We have examined the correlations between
inside-the-fibre $M_\star/L_z$ and fibre-colours and chosen the $M_\star/L_z$ versus $(r-z)$
colour as the best one (though other colours yield about equally good
correlations). Having established an empirical relation parametrised by
$M_\star/L_z = a (r-z) + b$ (which we did separately for passive and star-forming
galaxies), we can then, from an observed outside-the-fibre colour
estimate the corresponding outside-the-fibre $M_\star/L_z$.

\begin{figure}
\centerline{\includegraphics[scale=0.4]{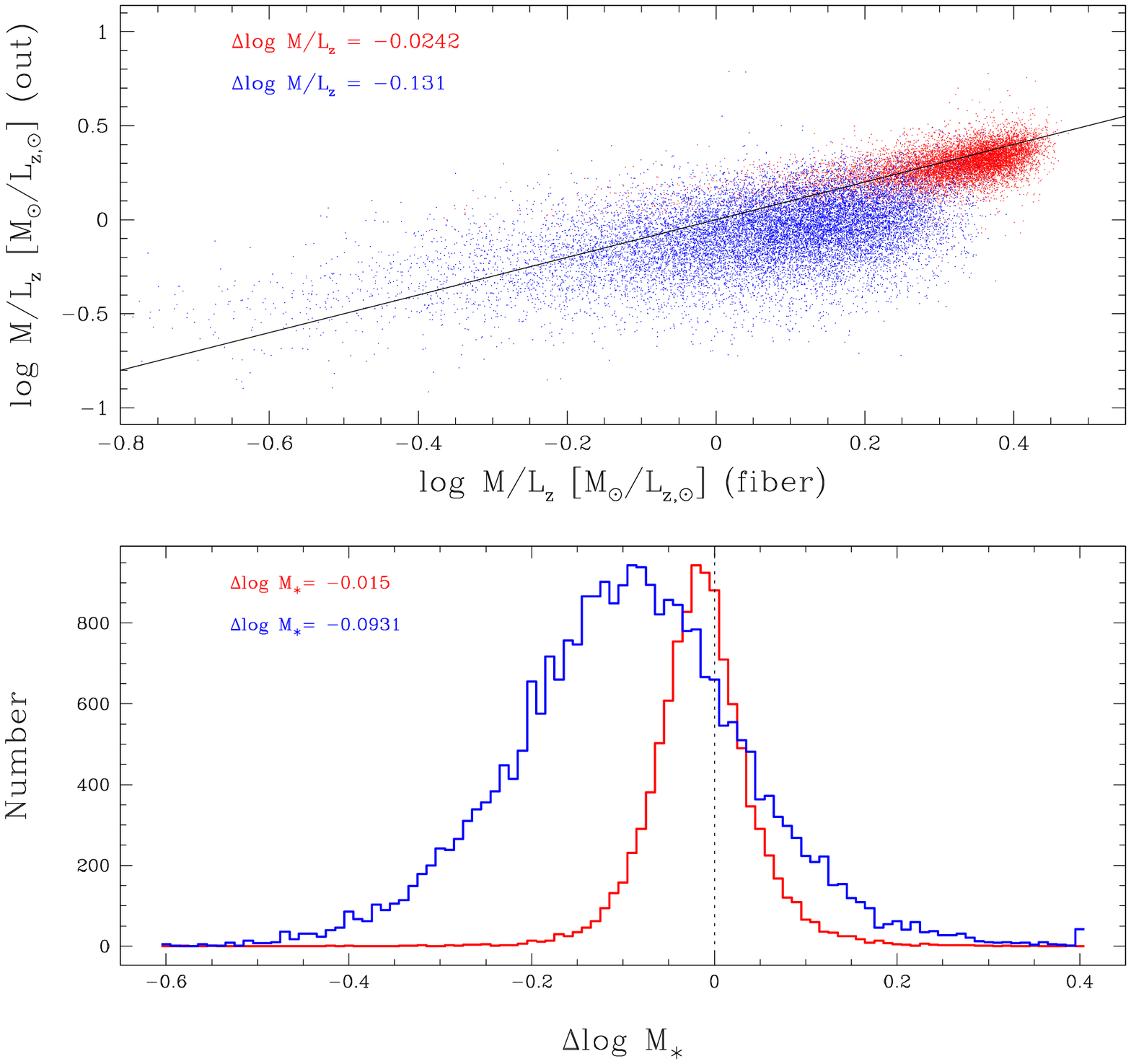}}
\caption{Top: relation between the mass-to-light ratio obtained inside the 3\arcsec\ fibre and outside it (see details in text) for star-forming galaxies (blue points) and passive galaxies (red points). Bottom: distribution of the differences between the total stellar masses estimated by considering a constant $M_\star/L_z$ along the radius of a galaxy and estimated via the outside-the-fibre $M_\star/L_z$, for star-forming galaxies (blue) and passive ones (red). }
\label{fig_appendix2}
\end{figure}

In the top panel of Fig.~\ref{fig_appendix2}, we show the relation between the $M_\star/L_z$ obtained inside the fibre and outside it, for star-forming and passive galaxies. Compared to the inside-the-fibre value, $M_\star/L_z$ is, in the median 0.13
dex smaller in the outside. This applies to star-forming
galaxies. This is expected, as the outside light is dominated by
disk-light, whereas the fibre data also includes the bulge.
Differences for the passive galaxies were negligible ($-0.03$ dex).

Total masses computed with this approximate correction are, in the
median, only 0.09 dex smaller than those obtained extrapolating the
$M_\star/L_z$ derived from the fibre-spectroscopic data to the whole galaxy, as can be seen in the bottom panel of Fig.~\ref{fig_appendix2}. Because this is a small difference, and also because the $M_\star/L_z$ versus colour calibration has a good deal of scatter, we prefer not to apply this correction in our analysis.

\label{lastpage}

\bsp

\end{document}